\documentclass[11pt]{article}
 
\usepackage{amsmath,amsthm}
\usepackage{amsthm}
\usepackage{amssymb}
\usepackage{amsfonts} 

\usepackage[latin1]{inputenc}

\usepackage{graphicx}

\def\be{\begin{equation}}
\def\ee{\end{equation}}
\def\bea{\begin{eqnarray}}
\def\eea{\end{eqnarray}}

\newcommand{\sect}[1]{\setcounter{equation}{0}\section{#1}}
\newcommand{\subsect}[1]{\subsection{#1}}
\newcommand{\subsubsect}[1]{\subsubsection{#1}}
\renewcommand{\theequation}{\arabic{section}.\arabic{equation}}

\newcommand{\Sk}{{\rm\ \!S}}            
\newcommand{\Ck}{{\rm\ \!C}}           
\newcommand{\Tk}{{\rm\ \!T}}             

  \newcommand{\kk}{\kappa}

 \def\1{\'{\i}}

\def\k{{\kappa}}

\newcommand{\te}{\phi}

\newcommand{\xx}{X}

\newcommand{\yy}{Y}

\newcommand{\uu}{u}

\newcommand{\ang}{J}

\newcommand{\mm}{{\cal E}_\k}

\newcommand{\mmb}{\epsilon}

\newcommand{\mmc}{\chi}

\newcommand{\mk}{|\k|}

\begin{document}

\

 \
 \smallskip
 
\vskip1cm

\begin{center}
 {\Large{\bf{The anisotropic oscillator on curved spaces: 
 \\[6pt] A new exactly solvable  model}}}

\bigskip

\bigskip

\begin{center}
{\sc \'Angel Ballesteros$^a$,      Francisco J.~Herranz$^a$, \c{S}engul Kuru$^b$ and Javier Negro$^c$}
\end{center}

\noindent
$^a$Departamento de F\1sica,  Universidad de Burgos,
09001 Burgos, Spain\\ ~~E-mail: angelb@ubu.es,  fjherranz@ubu.es\\

\noindent
$^b$Department of Physics, Faculty of Science,
 Ankara University, 06100 Ankara, Turkey\\
~~E-mail:  kuru@science.ankara.edu.tr
\\

\noindent
$^c$Departamento de F\'{\i}sica Te\'orica, At\'omica y
\'Optica, Universidad de Valladolid,\\  47011 Valladolid, Spain\\
~~E-mail:   jnegro@fta.uva.es

\end{center}

\medskip
\medskip
\medskip

\begin{abstract}
\noindent 
We present a new exactly solvable (classical and quantum) model that can be interpreted as the generalization to the two-dimensional sphere  and to the hyperbolic space  of the two-dimensional anisotropic oscillator with any pair of frequencies $\omega_x$ and $\omega_y$. The new curved Hamiltonian ${H}_\kappa$ depends on  the curvature $\kappa$ of the underlying space  as a deformation/contraction parameter, and the Liouville integrability of ${H}_\kappa$ relies on its separability in terms of geodesic parallel coordinates, which generalize the Cartesian coordinates of the plane.  Moreover, the system is   shown to be superintegrable  for   commensurate frequencies $\omega_x: \omega_y$, thus mimicking the behaviour of the flat Euclidean case, which is always recovered in the $\kappa\to 0$ limit.  The additional constant of motion in the commensurate case is, as expected, of higher-order  in the momenta and can be explicitly deduced by performing the classical factorization of the Hamiltonian. The known  $1:1$ and $2:1$ anisotropic curved oscillators  are recovered as particular cases of ${H}_\kappa$,  meanwhile all the remaining $\omega_x: \omega_y$ curved oscillators define new superintegrable systems. 
  Furthermore, the quantum Hamiltonian  $\hat {H}_\kappa$ is fully constructed and studied by following a quantum factorization approach. In the case of commensurate frequencies, the Hamiltonian $\hat {H}_\kappa$  turns out to be quantum superintegrable and leads to a new exactly solvable quantum model. Its corresponding spectrum, that exhibits  a maximal degeneracy, is explicitly given as an analytical deformation of the Euclidean eigenvalues in terms of both the curvature $\kappa$ and the Planck constant $\hbar$. In fact, such spectrum is obtained as a composition of two one-dimensional (either trigonometric or hyperbolic) P\"osch--Teller set of eigenvalues.
\end{abstract}

\bigskip\bigskip 


 \medskip


\newpage

%%%%%%%%%%%%%%%%%%%%%%%%%
\sect{Introduction}
The aim of this paper is to present a (classical and quantum) integrable generalization on the sphere $ {\mathbf S}^2$ and on  the hyperbolic plane $ {\mathbf H}^2$ of a unit mass two-dimensional anisotropic oscillator Hamiltonian
\be
{H} = \frac12(p_x^2 + p_y^2) + \frac{ 1}{2}( \omega_x^2  x^2 + \omega_y^2y^2) ,
\label{aa}
\ee
where $(x,y)\in \mathbb R^2$ are Cartesian coordinates, $(p_x,p_y)$  their conjugate momenta and the frequencies $(\omega_x,\omega_y)$ are arbitrary real numbers. 

It is well-known that the Euclidean  system~\eqref{aa} is always integrable (in the Liouville sense) due to its obvious separability in Cartesian coordinates. On the other hand, for  commensurate frequencies $\omega_x: \omega_y$ the Hamiltonian (\ref{aa}) defines a superintegrable oscillator, since an  ``additional" integral of motion of higher-order in the momenta arises (see~\cite{Jauch, Stefan, Winternitz} and references therein). We recall that in the classical case the superintegrability property ensures that all the bounded trajectories are closed, thus leading to Lissajous curves, while in the quantum case superintegrability gives rise to maximal degeneracy of the spectrum.

To the best of our knowledge,  
the only two known generalizations on the sphere and the hyperbolic space of the superintegrable anisotropic oscillator~\eqref{aa}  are the 1\,:\,1   and   2\,:\,1 cases.  In the classical case, both systems arise within the classification of superintegrable systems on $ {\mathbf S}^2$ and  $ {\mathbf H}^2$ that are endowed  with constants of motion that are {quadratic} in the momenta~(see \cite{mariano99,Kalnins1,Kalnins2}). In this paper we present the generalization of this result for arbitrary commensurate frequencies $\omega_x: \omega_y$, and the classical and quantum superintegrability of the proposed Hamiltonian ${H}_\k$,
where $\k$ stands for the curvature of the surface, will be proven by making use of a factorization approach. We remark that some preliminary results on the classical Hamiltonian ${H}_\kappa$  have recently been anticipated in~\cite{Procs}. The key point of our approach is that,  in the same way 
as Cartesian coordinates
are the natural ones to write the anisotropic oscillator in the Euclidean plane,
so are the geodesic parallel coordinates on constant curvature surfaces to express
the curved anisotropic oscillators. Moreover, in order to show that our systems consist in a deformation of the Euclidean
anisotropic oscillator we have adopted a notation depending explicitly
on the curvature parameter $\k$. For $\k>0$ we have the system defined
on the sphere, for $\k<0$ on the hyperboloid, while in the limit
$\k\to0$ we will recover all the well-known Euclidean results.

In Sections 2 and 3 we start by reviewing the integrability properties of classical and quantum anisotropic oscillators on the Euclidean plane $\mathbf{E}^2$ through the factorization approach 
introduced in~\cite{schrodinger40,infeldhull51,cooper94,oscar04,fernandez10,Kuru1} (see also~\cite{david,Kuru2,Kuru3} and references therein). In Section 4 we will revisit the well-known $1\!:\!1$ and $2\!:\!1$   curved oscillators~\cite{mariano99, ballesteros13, anisotropic}, and we will show that both Hamiltonians can be written in a simple and unified way if we express them in terms of the so called geodesic parallel coordinates on $ {\mathbf S}^2$ and  $ {\mathbf H}^2$ (see~\cite{mariano99,conf, Letter, Montreal}). This fact turns out to be the keystone for the generalization of the system in the case of arbitrary frequencies, which is presented in Section 5 and is shown to be superintegrable through the very same factorization approach that was used in the Euclidean case.

The quantization of the previous result is presented in Section 6, where the ladder and shift operators for  $\hat{H}_\kappa$ are explicitly constructed. From them, higher-order symmetries leading to the quantum superintegrability are straightforwardly obtained. Sections 7 and 8 are devoted to a detailed analysis of the spectral problem of the quantum commensurate oscillator on the sphere and the hyperbolic space, respectively. In particular, the maximal degeneracy of the energy levels will be explicitly shown. A final section including some remarks and open problems close the paper, and some technical tools or proofs that are needed along the paper have been included in the Appendixes.

%%%%%%%%%%%%%%%%%%%%%%%%%%%%%%%%%%%%%%%%%%%%%%%%%%%%%%%%%%%%%
\sect{The classical factorization method}

In this section we review the Hamiltonian~\eqref{aa} from the classical factorization viewpoint introduced in~\cite{Kuru1,david,Kuru2,Kuru3}. Although the results here presented are quite elementary, they are useful in order to present the approach that we will follow in the curved case. If we denote
\be
 \omega_x=\gamma \omega_y ,\qquad  \omega_y=\omega ,\qquad  \gamma\in  \mathbb R^+/\{0\},
 \label{ab}
\ee
the   Hamiltonian (\ref{aa}) can be  rewritten, in terms of the parameter $\gamma$ and the frequency $\omega$,  as
\begin{equation}
H = \frac12(p_x^2 + p_y^2) + \frac{\omega^2}{2} \left( (\gamma x)^2 + y^2 \right) .
\label{ac}
\end{equation}
Now, we introduce the new canonical variables
\be
\xi=\gamma x,\qquad  p_\xi =p_x/\gamma,\qquad \xi\in  \mathbb R,
\label{ad}
\ee
yielding
\begin{equation}
H= \frac12\, p_y^2 + \frac{\omega^2}{2} \,  y^2 
 +  \gamma^2\left(\frac12\, p_\xi^2 + \frac{\omega^2}{2\gamma^2} \,  \xi^2\right) .
 \label{ae}
\end{equation}
The two one-dimensional Hamiltonians $H^\xi$ and $H^y$ given by
\be
H^\xi = \frac12\, p_\xi^2 + \frac{\omega^2}{2\gamma^2} \xi^2  ,\qquad  H^y = \frac12\, p_y^2 + \frac{\omega^2}{2} y^2  ,\qquad H= H^y 
 +  \gamma^2 H^\xi , 
\label{af}
\ee
are indeed two integrals of the motion for $H$, since $\{ H , H^\xi \}=  \{ H , H^y \}=  \{ H^\xi , H^y \}=0$.

The factorization approach is based on the definition of the so-called ``ladder    functions" $B^\pm$, which are obtained by requiring that $H^\xi =B^+ B^-$, and  take the expression
\begin{equation}
B^{\pm} ={\mp}\frac{i}{\sqrt{2}}\, p_{\xi}+
\frac{1}{\sqrt2} \frac{\omega}{\gamma}\,  \xi \, .
\label{bc0}
\end{equation}
They lead to the Poisson algebra
\be
\{H^\xi ,B^{\pm} \}=\mp i\, \frac{\omega}{\gamma} \,B^{\pm} , \qquad  \{ B^-,B^+ \}= - i \,\frac{\omega}{\gamma} .
\label{commpt20}\nonumber
\ee
Therefore the functions $(H^\xi,B^\pm,1)$ generate the harmonic oscillator Poisson--Lie algebra $\mathfrak{h}_4$.

On the other hand, the ``shift functions'' $A^\pm$  also arise  by imposing that 
$H^y= A^+ A^-$, thus yielding
\begin{equation}
A^{\pm}=\mp \frac{i}{\sqrt{2}}\,p_{y}-\frac{\omega}{\sqrt{2}}\,  y \, .
\label{capm0}
\end{equation}
The four functions  $(H^\xi,A^\pm,1)$  span again the Poisson--Lie   algebra $\mathfrak{h}_4$, since
 \be
\{H^y ,A^{\pm} \}=\pm  i  {\omega}  A^{\pm} , \qquad  \{ A^-,A^+ \}=   i {\omega}  .
\label{ag}\nonumber
\ee
We stress that in this flat model the terms ``ladder'' and ``shift''
 are fully equivalent and could be interchanged. However, in the curved cases both sets of functions will  no longer be equivalent. 

Consequently, the Hamiltonian (\ref{ae}) can be rewritten in terms of the above ladder and shift functions as
\begin{equation}
H= A^+ A^- +\gamma^2 B^+ B^-,\qquad \{H ,B^{\pm}\}=\mp i {\gamma} {\omega} B^{\pm}   ,\qquad \{H ,A^{\pm}\}=\pm{ i \omega} A^{\pm}\,.
\label{commpt30}\nonumber
\end{equation}
The remarkable point now is that if we consider a rational value for $\gamma$,
\be
\gamma  = \frac{\omega_x}{\omega_y}=\frac{m}{n}   , \qquad m,n\in \mathbb N^\ast ,
\label{ah}
\ee
we obtain  two complex
constants of motion $X^{\pm}$ for $H$  (\ref{ae}) such that
\be
 \{H,X^{\pm}\}= 0\, ,
\label{ai}\nonumber
\ee
where we have defined
\begin{equation}
X^{\pm}=(B^{\pm})^n (A^{\pm})^{m}\, , \qquad \bar{X}^+= {X}^- ,  
\label{csymmet10}
\end{equation}
being $\bar{X}^+$ the complex conjugate of ${X}^+$.
The four constants of motion so obtained $(H^\xi,H^y,X^\pm)$,  are not functionally independent since from the factorization properties of 
$A^\pm$ and $B^\pm$ it can be proven that
\begin{equation}
X^+X^- = (H^\xi)^n(H^y)^m \,.
\nonumber
\end{equation}
In fact, these four functions generate a polynomial Poisson algebra, namely,
\bea
&&  \{H^\xi,X^{\pm}\}= \mp i\omega\, \frac{n^2}m X^\pm, \qquad   \{H^y,X^{\pm}\}= \pm i\omega \, m X^\pm , \nonumber\\[2pt]
&& \{ X^+,X^-\}= - i\, \frac{\omega }{m}(H^\xi)^{n-1} (H^y)^{m-1} \left( 
  {m^3 H^\xi- n^3 H^y}  \right) .
\label{aii}
\eea

Notice that the integrals of motion (\ref{csymmet10}) are of $(m+n)$th-order in the momenta and, since $X^{\pm}$ are complex functions we can get real constants of motion given by
\be
\xx= \frac 12(X^+ + X^-),\qquad \yy= \frac 1{2i} (X^+ - X^-).
\label{aj}
\ee
The degree in the momenta for one of them is $(m+n)$ 
and $(m+n-1)$ for the other one.  From (\ref{aii}) we find the following  polynomial algebra of real symmetries
\bea
&& \{ H^\xi,\xx\}=\omega\,\frac{n^2}m \yy ,\qquad \  \{ H^\xi,\yy\}=-\omega\,\frac{n^2}m \xx ,\nonumber\\[2pt]
&& \{ H^y,\xx\}=-\omega\, m \yy ,\qquad \{ H^y,\yy\}=\omega\, m \xx ,\nonumber\\[2pt]
&&\{\xx,\yy\}= \frac{ \omega}{2m} (H^\xi)^{n-1} (H^y)^{m-1} \left( 
  {m^3 H^\xi- n^3 H^y}  \right) ,
\label{ak} 
\eea
which is generated by the four real integrals $(H^\xi,H^y,X,Y)$.

In this way, we have recovered all the well-known results on the (super)integrability of anisotropic oscillators~\cite{Jauch, Stefan, Winternitz} which   can be summarized as follows.

\medskip

\noindent
{\bf Theorem 1.} {\em {\rm (i)} The Hamiltonian $H$ (\ref{ae}) is integrable for any value of the real parameter $\gamma$, since it is endowed with a quadratic constant of motion given by either $H^\xi$ or $H^y$ (\ref{af}).

\noindent
 {\rm (ii)} When $\gamma=m/n$ is a rational parameter (\ref{ah}), the Hamiltonian (\ref{ae}) defines a  superintegrable  anisotropic oscillator with commensurate frequencies $\omega_x:\omega_y$ and the additional constant of motion is given by either $X$ or $Y$ in (\ref{aj}). The sets $(H,H^\xi,\xx)$ and $(H,H^\xi,\yy)$ are formed by three functionally independent functions.}

 %%%%%%%%%%%%%%%%%%%%%%%%%
\subsect{The  1\,:\,1  oscillator}

   If $\gamma=1$, we can set $m=n=1$ such that $\omega_x=\omega_y=\omega$ and (\ref{ad}) gives $\xi=x$ and $p_\xi=p_x $. Hence, we recover the isotropic oscillator 
   $$
{H}^{1:1} = \frac12(p_x^2 + p_y^2) + \frac{ \omega^2 }{2}( x^2 + y^2) ,
  $$
  and the integrals (\ref{aj}) reduce to
\be
\xx=-\tfrac 12 (p_x p_y +\omega^2 x y) ,\qquad \yy =-\tfrac 12 \omega (x p_y - y p_x)   .
\label{al}
\ee
Therefore, the quadratic integral $\xx$  is one of the components of the Demkov--Fradkin tensor~\cite{Demkov,Fradkin}, meanwhile $\yy$ is proportional to the angular momentum 
\be
\ang = x p_y - y p_x .
\label{all}
\ee
Since $m+n$ is even, the symmetry with highest degree is $X$ 
and the lowest one is given by $Y$.

 %%%%%%%%%%%%%%%%%%%%%%%%%

\subsect{The   2\,:\,1  oscillator}

For $\gamma=2$, we take $m=2$ and $n=1$. Thus $\omega_x=2\omega_y=2\omega$, $\xi=2 x$ and $p_\xi =p_x/2$. The    Hamiltonian (\ref{ae})  and the  integrals (\ref{aj}) read 
\bea
&& H^{2:1} = \frac12\, p_y^2 + \frac{\omega^2}{2} \, y^2 
 +  4\left(\frac12\, p_\xi^2 + \frac{\omega^2}{8}\, \xi^2\right) =\frac12(p_x^2 + p_y^2) + \frac{\omega^2}{2} \left( 4 x^2 + y^2 \right) ,\nonumber\\[2pt]
 && \xx=-\frac{\omega}{4\sqrt{2}} \left( p_y ( \xi p_y - 4 y p_\xi) - \omega^2 \xi y^2  \right)= -\frac{\omega}{2\sqrt{2}} \left( p_y \ang - \omega^2 x y^2  \right) ,\nonumber\\[2pt]
 &&\yy=  \frac{1}{2\sqrt{2}} \left( p_\xi p_y^2 +   \omega^2 y ( \xi p_y -   y p_\xi)    \right)=  \frac{1}{4\sqrt{2}} \left( p_x p_y^2 +   \omega^2 y ( 4 x p_y -   y p_x)    \right) .
\label{am}
\eea
The quadratic symmetry $\xx$, which  involves the angular momentum $\ang$ (\ref{all}), is the constant considered in the literature (see e.g.~\cite{mariano99,WolfBoyer}), whilst    $\yy$ is a cubic integral. In this sense, the $2:1$ oscillator can be considered as a superintegrable system with {\em quadratic} constants of motion. In fact,   the $1:1$ and $2:1$ oscillators are the only anisotropic Euclidean oscillators endowed with quadratic integrals  (see the classifications~\cite{mariano99,evans}), and all the remaining $m:n$ oscillators have higher-order symmetries. In this case, since $m+n$ is odd, the highest $(m+n)$-degree symmetry is $Y$ while  $X$ is
 of lowest order $(m+n-1)$.

 %%%%%%%%%%%%%%%%%%%%%%%%%

\subsect{The  1\,:\,3  oscillator}

In this case we have that  $\gamma=1/3$, $m=1$,  $n=3$,  $\omega_x= \omega_y/3= \omega/3$, $\xi=x/3$ and $p_\xi =3 p_x$. The    Hamiltonian (\ref{ae}) is now given by
$$
 H^{1:3}= \frac12\, p_y^2 + \frac{\omega^2}{2} y^2 
 +  \frac 19\left(\frac12\, p_\xi^2 + \frac{9\omega^2}{2} \, \xi^2\right) =\frac12(p_x^2 + p_y^2) + \frac{\omega^2}{2} \left( \frac 19 \, x^2 + y^2 \right) ,
$$
and the  integrals (\ref{aj}) are
\bea
 && \xx= 
 \frac{1}{4 } \left( 27 p_x^3 p_y           - 9 \omega^2 x p_x(   x p_y -   3 y p_x) -  \omega^4 x^3 y  \right) ,\nonumber\\[2pt]
 &&\yy=  
  \frac{\omega}{4} \left( 27 p_x^2  \ang  -    \omega^2 x^2 ( x p_y -   9 y p_x)    \right).
\label{an}\nonumber
\eea
Since the symmetry $\xx$ is   quartic   in the momenta but $\yy$ (that includes the    angular momentum   (\ref{all})) is cubic, $H$ is a cubic superintegrable system.

Notice that, obviously, the $1:2$ and $3:1$ oscillators with $\gamma=1/2$  and $\gamma=3$ define   equivalent systems to the previous oscillators  via the interchange $x\leftrightarrow y$.  And, clearly, 
any    $m:n$  oscillator  (with $\gamma$) is equivalent to the  $n:m$  one (with $1/\gamma$). Surprisingly enough, this (trivial) fact from the Euclidean viewpoint will no longer hold when the curvature of the space is non-vanishing, as we will explicitly show in Section 5.

%%%%%%%%%%%%%%%%%%%%%%%%%%%%%%%%%%%%%%%%%%%%%%%%%%%%%%%%%%%%%
\sect{The quantum factorization method}

In order to study the quantum analog of the Hamiltonian (\ref{aa}), let us introduce the standard definitions for   quantum position $(\hat x,\hat y)$ and  momentum $(\hat p_x,\hat p_y)$ operators  
\be
\hat x\Psi(x,y) = x \Psi(x,y),\qquad \hat p_x\Psi(x,y) = - i \hbar  \, \frac{\partial \Psi(x,y)}{\partial x},
\qquad [\hat x, \hat p_x]= i\hbar   ,
\label{ba}\nonumber
\ee
and similarly for $(\hat y,{\hat p}_y)$. Hereafter, the hat will be suppressed 
for the  $\hat x,\hat y$ position operators to simplify the presentation.
Hence, as it is well-known, the quantum version of the  Hamiltonian  (\ref{aa}) reads
\be
{\hat H} = \frac12(\hat p_x^2 + \hat p_y^2) + \frac{ 1}{2}( \omega_x^2   x^2 + \omega_y^2  y^2) 
=-   \frac{\hbar^2}2 \left( \frac{\partial^2 }{\partial x^2} + \frac{\partial^2 }{\partial y^2}  \right) 
+ \frac{ 1}{2}( \omega_x^2  x^2 + \omega_y^2 y^2)  .
\label{bb}\nonumber
\ee
By introducing the frequency $\omega$ (\ref{ab}) and the new variable $\xi$ (\ref{ad})  we find   that
 \begin{equation}
\hat{H} = -\frac{\hbar^2}{2} \, \frac{ \partial^2}{\partial\,y^2}
+\frac{\omega^2}{2} \,  y^2 
+\gamma^2 \left(-\frac{\hbar^2}{2} \, \frac{ \partial^2}{\partial\,\xi^2}+ \frac{\omega^2}{2 \gamma^2}\, \xi^2
\right) ,
\label{hq3p2}
\end{equation}
and the corresponding eigenvalue equation is given by   
\begin{equation}\label{eve1a}\nonumber
\hat{H}\,\Psi(\xi,y)=E \,\Psi(\xi,y)\,.
\end{equation}
From (\ref{hq3p2})   we   get  the one-dimensional   Hamiltonian operators
\begin{equation}\label{hqyp}
\hat{H}^{\xi} =-\frac{\hbar^2}{2} \, \frac{ \partial^2}{\partial\,\xi^2}+ \frac{\omega^2}{2 \gamma^2}\, \xi^2 ,\qquad 
\hat{H}^{y} =  -\frac{\hbar^2}{2} \, \frac{ \partial^2}{\partial\,y^2}+\frac{\omega^2}{2} \,  y^2  ,\qquad 
 {\hat H} ={\hat H}^{y} +\gamma^2 {\hat H}^\xi  \, , 
\end{equation}
such that $[ {\hat H} ,{\hat H}^{\xi} ]=[ {\hat H} ,{\hat H}^{y} ]=[ {\hat H}^{\xi}  ,{\hat H}^{y} ]=0$.

Now, we look for factorized solutions,
$\Psi(\xi,y)=\Xi(\xi)\,Y(y)$, where the component
functions $\Xi(\xi)$ and $Y(y)$  satisfy the following one-dimensional eigenvalue equations
\begin{equation}\label{eve2a}\nonumber
\hat{H}^{\xi}\,\Xi(\xi)=E^{\xi}\,\Xi(\xi),
\qquad\hat{H}^{y}\,Y(y)=E^{y}\,Y(y) .
\end{equation}
The factorizations of these systems are the standard ones
\begin{equation}\label{fachxyp}
\hat{H}^\xi=\hat{B}^+\hat{B}^-+\lambda^B, \qquad 
\hat{H}^y=\hat{A}^+\hat{A}^-+\lambda^A ,
\end{equation}
and yield the following ladder  
\begin{equation}\nonumber
\hat{B}^{\pm}=\mp \frac{\hbar}{\sqrt{2}} \, \frac{ \partial}{\partial \xi} + 
\frac{\omega}{\sqrt{2}\gamma}\,\xi,\qquad 
\lambda^{B}= \frac{\hbar\omega}{2 \gamma} ,
\end{equation}
and shift operators
\begin{equation}\nonumber
\hat{A}^{\pm}=\mp \frac{\hbar}{\sqrt{2}} \, \frac{ \partial}{\partial y}- \frac{\omega}{\sqrt{2}}\,y,\qquad \lambda^{A}= -\frac{\hbar\omega}{2}\,.
\end{equation}
The commutation rules of the two sets of operators $(\hat H^\xi, \hat B^\pm)$ and $(\hat H^y, \hat A^\pm)$ read
\bea
&& [\hat H^\xi,\hat{B}^{\pm}] = \pm\frac{\hbar \omega}{ \gamma}  \hat{B}^{\pm} , \qquad [\hat B^-,\hat B^+]=\frac{\hbar\omega}{\gamma} ,
\nonumber\\[2pt]
&&
[\hat H^y,\hat{A}^{\pm}] = \mp  \hbar \omega \hat{A}^{\pm} ,\qquad [\hat A^-,\hat A^+]=-{\hbar\omega} ,
\nonumber
\eea
and  each set generates   the harmonic oscillator Lie algebra $\mathfrak{h}_4$. Hence, we find that
\be
\hat{H}^\xi=\hat{B}^+\hat{B}^-+ \frac{\hbar\omega}{2 \gamma} =\hat{B}^-\hat{B}^+- \frac{\hbar\omega}{2 \gamma}   , \qquad 
\hat{H}^y=\hat{A}^+\hat{A}^-    -\frac{\hbar\omega}{2} =\hat{A}^-\hat{A}^+     +\frac{\hbar\omega}{2}   .
\label{be}
\ee
 
The eigenvalues of $\hat{H}^\xi$ and  $\hat{H}^y$ corresponding
to the respective eigenfunctions $\Xi^\mu(\xi)$ and $Y^\nu(y)$ turn out to be
\begin{equation}\label{eigenvaluexyp}
{E}^{\xi,\mu}=  \frac{\hbar\omega}{2 \gamma}+  \mu\,   \frac{\hbar\omega}{ \gamma} ,
\qquad {E}^{y,\nu}=\frac{\hbar\omega}{2}+\nu\,    {\hbar\omega} ,\qquad \mu,\nu=0,1,2,\dots
\end{equation}
Next, according to (\ref{hqyp}) and  (\ref{be})    
   the  Hamiltonian (\ref{hq3p2})  can be written as 
\be
\hat H= 
\frac 12\bigl( \hat A^+  \hat A^- + \hat A^-  \hat A^+ \bigr) +\frac{\gamma^2}2\bigl( \hat B^+  \hat B^- + \hat B^-  \hat B^+  \bigr) ,
\label{bc1}\nonumber
\ee
where
\begin{equation}
[\hat H,\hat{B}^{\pm}] = \pm\gamma \hbar  \omega \hat{B}^{\pm} , \qquad
[\hat H,\hat{A}^{\pm}] = \mp  \hbar \omega \hat{A}^{\pm}   .
\nonumber
\end{equation}
Finally, the eigenvalue of the wave function
$\Psi^{\mu,\nu}(\xi,y) =\Xi^\mu(\xi)Y^\nu(y)$, corresponding to   (\ref{eigenvaluexyp})  reads  
\begin{equation}\label{eigenvaluep}
{E}^{\mu,\nu}={E}^{y,\nu}+\gamma^2   {E}^{\xi,\mu}=\hbar \omega\left( \tfrac 12 (\gamma+1) +\gamma \mu+\nu    \right) ,\qquad \mu,\nu=0,1,2,\dots
 \end{equation}
 
 Furthermore,  if $\gamma =m/n$, with  $m,n\in \mathbb N^\ast$, as in   the classical case,  we obtain ``additional" higher-order symmetries for $\hat H$ (\ref{hq3p2}), beyond $\hat H^\xi $  and $\hat H^y$, since the operators
\begin{equation}\label{symmpm2p}
\hat{X}^{\pm}=(\hat{B}^{\pm})^n\, (\hat{A}^{\pm})^{m} \qquad
\mbox{are such that} \qquad [\hat H,\hat X^\pm]=0 \,.
\end{equation}
From the factorization properties (\ref{fachxyp}) 
of $\hat A^\pm$ and $\hat B^\pm$ we find that
\begin{equation}\nonumber
\hat X^+\hat X^- = 
\left(\hat H^\xi- \frac{\hbar \omega}{2\gamma} \right)^n \left(\hat H^y+ \frac{\hbar \omega}{2} \right)^m \, .
\end{equation}
Therefore, the four constants of motion 
$(\hat H^y, \hat H^\xi, \hat X^\pm)$ are algebraically dependent, as they should be.
On the other hand, the sets $(\hat H^y, \hat H^\xi, \hat X^+)$
or $(\hat H^y, \hat H^\xi, \hat X^-)$ are algebraically independent.
Notice that, obviously, we could also consider the quantum observables $\hat \xx$ and $\hat\yy$ defined by following (\ref{aj}) and that the sets
$(\hat H^y, \hat H^\xi, \hat X^\pm)$ or $(\hat H^y, \hat H^\xi, \hat X, \hat Y)$
close a polynomial symmetry algebra similar to (\ref{aii}) or (\ref{ak}).

As a straightforward consequence, if  $\gamma=m/n$ is a rational parameter (\ref{ah}),   the energy levels of $\hat H$  can be  degenerate, since~\eqref{eigenvaluep} can be written in the form
 \begin{equation}\nonumber
    {E}^{\mu,\nu}=\hbar \omega\left( \frac 12 \left(\frac{m}{n}+1\right) +\frac{m\, \mu+ n\, \nu}{n}  \right) ,\qquad \mu,\nu=0,1,2,\dots
 \end{equation}
and the energy will be the same for all pairs $(\mu,\nu)$ of quantum numbers for which $m\, \mu+ n\, \nu$ takes the same value. Moreover, the corresponding eigenstates
are connected by means of the $\hat X^\pm$ operators.

 Summarizing, the quantum counterpart of Theorem 1 can be stated as follows.

\medskip

\noindent
{\bf Theorem 2.} {\em {\rm (i)} The Hamiltonian $\hat H$    (\ref{hq3p2}) commutes with the operators $\hat H^\xi$ and $\hat H^y$ (\ref{hqyp}) and
defines   an   integrable quantum  system for any value of the real parameter $\gamma$. The  discrete spectrum of $\hat H$ depends on two quantum numbers and is given by  $E^{\mu,\nu}$ (\ref{eigenvaluep}).

\noindent
 {\rm (ii)} Whenever $\gamma=m/n$ is a rational parameter, the Hamiltonian  $\hat H$   commutes with the observables $\hat{X}^{\pm}$ (\ref{symmpm2p}). The   sets $(\hat H,\hat H^\xi, \hat X^+)$ and  $(\hat H,\hat H^\xi,\hat X^-)$ are formed by three algebraically independent observables.
Therefore,  the quantum anisotropic oscillator with commensurate frequencies $\omega_x:\omega_y$  is a superintegrable quantum model,   and the spectrum $E^{\mu,\nu}$ is degenerate.}

%%%%%%%%%%%%%%%%%%%%%%%%%%%%%
\sect{The (known) 1\,:\,1   and   2\,:\,1 curved oscillators}

The $1\!:\!1$ and $2\!:\!1$ cases are the only superintegrable Euclidean oscillators  
with integrals
of motion that are quadratic in the momenta, and they are the only ones whose corresponding curved generalizations are well-known. In this section we will recall both of them, and we will show that by rewriting these two Hamiltonians in terms of geodesic parallel coordinates we will find the keystone in order to propose the  curved analogue of the generic $m:n$ oscillator.

The isotropic 1\,:\,1 oscillator is usually expressed in geodesic polar coordinates since in this way the isotropic oscillator potential on $ {\mathbf S}^2$  (characterized by the curvature $\k=1)$ and $ {\mathbf H}^2$  (with $\k=-1)$ is simply written in 
terms of the functions $\tan^2 r$ and $\tanh^2 r$, respectively, where the variable $r$ is just the geodesic distance from the particle to the centre of force. In this way, the first-order expansion for both potentials around $r=0$ gives the Euclidean potential function $r^2$. This system is also known as the Higgs oscillator~\cite{Higgs,Leemon} and has been widely studied 
(see~\cite{mariano99, ballesteros13,Pogoa,
Nersessian1,Santander6,Nersessian2,
Annals09} and references therein).

In order to be able to consider simultaneously the two curved spaces, 
and  to take the Euclidean limit as the zero curvature case, 
throughout the paper we will make use of the $\k$-dependent cosine and sine functions defined  
by 
\begin{equation}
\Ck_{\k}(\uu)\equiv\sum_{l=0}^{\infty}(-\k)^l\frac{\uu^{2l}} 
{(2l)!}=\left\{
\begin{array}{ll}
  \cos {\sqrt{\k}\, \uu} &\quad  \k>0 \\ 
\qquad 1  &\quad
  \k=0 \\
\cosh {\sqrt{-\k}\, \uu} &\quad   \k<0 
\end{array}\right.  ,
\nonumber
\end{equation}
\begin{equation}
   \Sk{_\k}(\uu) \equiv\sum_{l=0}^{\infty}(-\k)^l\frac{\uu^{2l+1}}{ (2l+1)!}
= \left\{
\begin{array}{ll}
  \frac{1}{\sqrt{\k}} \sin {\sqrt{\k}\, \uu} &\quad  \k>0 \\ 
\qquad \uu  &\quad
  \k=0 \\ 
\frac{1}{\sqrt{-\k}} \sinh {\sqrt{-\k}\, \uu} &\quad  \k<0 
\end{array}\right.  .
\nonumber
\end{equation}
Obviously, the $\k$-tangent is defined
by 
$$
 \Tk_\k(\uu)\equiv\frac{\Sk_\k(\uu)}{\Ck_\k(\uu)} .
 $$
Some  relations  involving these $\k$-functions   can be found in~\cite{mariano99,conf} and,  more  extensively,   in~\cite{trigo}. For instance: 
\bea
&&\Ck^2_\k(\uu)+\k\Sk^2_\k(\uu)=1,\qquad  \Ck_\k(2\uu)= \Ck^2_\k(\uu)-\k\Sk^2_\k(\uu), \qquad \Sk_\k(2\uu)= 2 \Sk_\k(\uu) \Ck_\k(\uu) , \nonumber\\[2pt]
&& \frac{ {\rm d}}{{\rm d} \uu}\Ck_\k(\uu)=-\k\Sk_\k(\uu),\qquad         \frac{ {\rm d}}
{{\rm d} \uu}\Sk_\k(\uu)= \Ck_\k(\uu)  ,\qquad 
\frac{ {\rm d}}
{{\rm d} \uu}\Tk_\k(\uu)=  \frac{1}{\Ck^2_\k(\uu) }.
\label{za}
\eea

In terms of the curvature $\k$ and the geodesic polar  coordinates $(r,\phi)$ the complete  Higgs Hamiltonian is given by (see, for instance,~\cite{ballesteros13})
\be
H_\k^{1:1}={\cal T}_\k+ U_\k^{1:1}= \frac 12 \left( p_r^2+\frac{p_\te^2}{\Sk^2_\kk(r)} \right)+ \frac{\omega^2}2  \Tk^2_\k(r),
\label{da}
\ee
and the specific expressions for  ${\mathbf S}^2$  ($\k=1)$ and $ {\mathbf H}^2$  ($\k=-1)$ are straightforwardly obtained.

On the other hand, the superintegrable 2\,:\,1 curved oscillator,  with Hamiltonian $H_\k^{2:1}={\cal T}_\k+ U_\k^{2:1}$, was firstly introduced in the classification carried out in~\cite{mariano99} and it has been recently studied in detail in~\cite{ballesteros13, anisotropic}. In geodesic polar variables the potential $U_\k^{2:1}$ adopts the (rather cumbersome) expression
  \be 
U_\k^{2:1} = \frac{\omega^2}{2} \left( \frac{4\Tk^2_\kk(r)\cos^2\te}{\left(1-\kk \Sk^2_\kk(r)\sin^2\te\right)\left(1-\kk \Tk^2_\kk(r)\cos^2\te\right)^2} +  \frac{\Sk^2_\kk(r)\sin^2\te}{ 1-\kk \Sk^2_\kk(r)\sin^2\te }     \right)  .
    \label{dc}
  \ee
Therefore, a glimpse on~\eqref{da} and~\eqref{dc} makes evident that the generalization of these potentials for the arbitrary $m : n$ case is far from being obvious.

%%%%%%%%%%%%%%%%%%%% figure1 %%%%%%%%%%%%%%%%%%%%%%%

\begin{figure}[t]
\begin{center}
\begin{picture}(170,145)
\footnotesize{
\put(127,22){$\bullet$}
\put(52,33){\makebox(0,0){$\te$}}
 \qbezier(46,25)(46,35)(40,38)
 \put(80,113){\makebox(0,0){$x'$}}
\put(17,62){\makebox(0,0){$y'$}}
\put(133,62){\makebox(0,0){$y$}}
\put(80,33){\makebox(0,0){$x$}}
\put(75,74){\makebox(0,0){$r$}}
\put(142,112){\vector(4,3){1}}
\put(117,94){$\bullet$}
\put(138,15){\makebox(0,0){$P_1$}}
\put(137,99){\makebox(0,0){$P$}}
\put(149,80){\makebox(0,0){$l'_1$}}
\put(149,119){\makebox(0,0){$l$}}
\put(22,22){$\bullet$}
\put(15,15){\makebox(0,0){$O$}}
\put(22,105){$\bullet$}
\put(15,108){\makebox(0,0){$P_2$}}
\put(15,130){\makebox(0,0){$l_2$}}
\put(25,10){\vector(0,1){125}}
\put(0,25){\vector(1,0){170}}
\put(168,15){\makebox(0,0){$l_1$}}
\put(35,98){\line(0,1){10}}
\put(25,98){\line(1,0){10}}
\put(120,25){\line(0,1){10}}
\put(120,35){\line(1,0){9}}
\qbezier[50](25,108)(70,110)(140,90)
\qbezier[50](130,25)(125,80)(115,114)
\put(115,124){\makebox(0,0){$l'_2$}}
\qbezier(25,24)(50,50)(140,111)
}
\end{picture}
\end{center}
\noindent
\\[-40pt]
\caption{\footnotesize Schematic representation of the geodesic   coordinates $(x,y)$, $(x',y')$  and  $(r,\te)$  of a   point  $P$   on a    curved space.}
\label{figure1}
\end{figure}
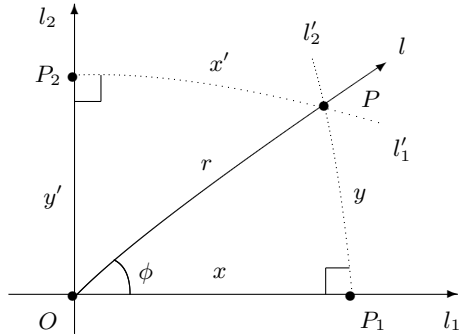

%%%%%%%%%%%%%%%%%%%%%%%%%%%%%%%%%%%%%%%%%%%

However, things drastically change  if we make use of the geodesic parallel coordinates for $ {\mathbf S}^2$  and $ {\mathbf H}^2$, that are described in detail in the Appendix A. In order to define them, we take an origin $O$ in the space,  two base geodesics $l_1$, $l_2$ orthogonal at $O$ and  the geodesic $l$ that joins a point $P$ (the particle) and $O$ (see Fig.~1). The {geodesic polar coordinates} $(r,\phi)$ are defined by the   distance $r$  between $O$ and $P$ measured  along $l$ and the   angle $\phi$ of $l$ relative to $l_1$. Let $P_1$ be the intersection point of $l_1$ with its orthogonal geodesic $l_2'$ through $P$. Then, the {geodesic parallel coordinates} $(x,y)$ are just defined by the distance $x$  between $O$ and $P_1$ measured  along $l_1$  and  the distance $y$  between $P_1$ and $P$ measured  along $l_2'$.  Notice that a similar set of coordinates $(x',y')$ can also be formed by considering   
   the intersection point  $P_2$ of $l_2$ with its orthogonal geodesic $l_1'$ through $P$ and  that generally $(x,y)\ne (x',y')$ if $\k\ne 0$ (see~\cite{mariano99,conf}). It is straightforward to realize that on 
 $ {\mathbf E}^2$ with $\k=0$, the coordinates $(x,y)=(x',y')$ reduce to Cartesian coordinates and $(r,\phi)$ give the usual polar ones.

In fact, the geodesic parallel coordinates turn out to be the closest to the Cartesian ones on these two curved spaces, and they will be indeed the appropriate ones in order to write the curved anisotropic oscillator Hamiltonians. This statement can be made evident if we consider the known curved $1:1$ and $2:1$ cases. Namely, if we  apply  to (\ref{da})  the relations (\ref{cf}) and the first identity given in (\ref{za}), we obtain 
two equivalent forms for  the potential $U_\k^{1:1}$ written in  geodesic parallel coordinates (see~\cite{mariano99}):
\be
U_\k^{1:1}= \frac{\omega^2}2 \left(\Tk^2_\k(x)+\frac{\Tk^2_\k(y)}{\Ck^2_\k(x)}  \right) = \frac{\omega^2}2 \left(\frac{\Tk^2_\k(x)}{\Ck^2_\k(y)} +\Tk^2_\k(y) \right) .\label{db}
\ee
By using the same transformation, the potential $U_\k^{2:1} $  (\ref{dc}) is shown to take the following expression in terms of geodesic parallel coordinates   
\be 
U_\k^{2:1} = \frac{\omega^2}{2} \left( \frac{ \Tk^2_\kk(2x) }{  \Ck^2_\kk(y) } +   \Tk^2_\kk(y)  \right)  .
    \label{dd}
  \ee
As a consequence, from both expressions $U_\k^{1:1}$ (\ref{db}) and $U_\k^{2:1}$ (\ref{dd})  it is natural to propose the following expression for the generic
curved anisotropic oscillator potential:
\be 
U_\k^{\gamma} = \frac{\omega^2}{2} \left( \frac{ \Tk^2_\kk(\gamma x) }{  \Ck^2_\kk(y) } +   \Tk^2_\kk(y)  \right)  ,\qquad \gamma\in    \mathbb R^+/\{0\}.
    \label{de}
    \ee
 In the next two sections, we will  show that this Ansatz is correct, since~\eqref{de} provides an integrable system both in the classical and quantum contexts, that can be exactly solved by making use of the factorization approach in terms of geodesic parallel coordinates. Moreover, in the commensurate case $\gamma=m/n$ the system turns out to be superintegrable due to the existence of an additional symmetry. We also remark that, generically, $U_\k^{\gamma} $ and $U_\k^{1/\gamma} $ will define two different  systems, in contradistinction with what happens in the Euclidean case.

%%%%%%%%%%%%%%%%%%%%%%%%%%%%%
\sect{The generic classical curved anisotropic oscillator}

Let us consider the following Hamiltonian function of a particle with unit mass written  in terms of geodesic parallel 
coordinates $(x,y)$
\begin{equation}\label{hc1}
{H}_{\kappa} =  {\cal T}_\k + {U}^\gamma_\k = \frac{1}{2} \left(\frac{p_{x}^2}{ \Ck_\k^2(y)}+p_{y}^2\right)+\frac{\omega^2}{2} 
\left(\frac{ \Tk_{\kappa}^2(\gamma x)}{ \Ck_{\kappa}^2(y)}+\Tk_{\kappa}^2(y)\right), 
\end{equation}
where   $\omega$ and  $\gamma$ are positive real constants, and the potential term is just~\eqref{de}.
Obviously, this Hamiltonian for $\k=0$ reproduces the Euclidean
anisotropic oscillator (\ref{ac}), while for $\k>0$ gives a system defined on  
$ {\mathbf S}^2$   and for $\k<0$ on $ {\mathbf H}^2$.
For $\kappa< 0$,
the Hamiltonian  is well defined for  any real values of $x$ and  $y$ (\ref{zb}). However, if $\k>0$ in order to avoid a multivalued Hamiltonian,  restrictions on the domain of the coordinates (\ref{zzbb}) arise from the term  $\Tk_{\kappa}(\gamma x)$,  namely:
\bea
 {\mathbf S}^2\ (\k>0) :&&     -\frac{\pi}{2\sqrt{\k}}< \gamma x< \frac{\pi}{2\sqrt{\k}} ,\quad   -\frac{\pi}{2\sqrt{\k}}< y< \frac{\pi}{2\sqrt{\k}} ,  \quad \gamma\ge \frac 12  . \label{rangeS}\\
 {\mathbf H}^2\ (\k<0) :&&  x,y \in \mathbb{R} ,\quad \gamma\in  \mathbb R^+/\{0\}.
 \label{range}
\eea

By assuming that $\k\ne  0$ and by using the relation
\be
 1+\k  \Tk_{\kappa}^2(u)=1/\Ck_{\kappa}^2(u),
\label{prop}
\ee
 which can be derived from (\ref{za}),     the  Hamiltonian ${H}_{\kappa}$ can
  be rewritten as 
\begin{equation}\label{hc2}
H_{\kappa} = 
\frac{p_{y}^2}{2}+\frac{1}{\Ck_{\kappa}^2(y)}\left(\frac{p_{x}^2}{2}+
\frac{\omega^2}{2 \kappa \Ck_{\kappa}^2(\gamma x)}\right)-
\frac{\omega^2}{2\kappa} ,\quad \k\ne 0.
\end{equation}
After introducing the new variable $\xi=\gamma x$
     (\ref{ad}) with domain given by~\eqref{rangeS} and~\eqref{range},     
   the Hamiltonian (\ref{hc2}) takes the   form 
\begin{equation}
H_{\kappa} = \frac{p_{y}^2}{2}+\frac{\gamma^2}{\Ck_{\kappa}^2(y)}\left(\frac{p_{\xi}^2}{2}+\frac{\omega^2}{2 \kappa  \gamma^2 \Ck_{\kappa}^2(\xi)}\right)-
\frac{\omega^2}{2  \kappa}\, ,\qquad \k\ne 0.
\nonumber
\end{equation}
In this way the total Hamiltonian can be rewritten as 
\begin{equation}\label{hc3}
H_{\kappa}  = \frac{p_{y}^2}{2}+\frac{\gamma^2 H_{\kappa}^\xi}{\Ck_{\kappa}^2(y)}-
\frac{\omega^2}{2\kappa} ,
\end{equation}
where the constant of the motion $H_{\kappa}^\xi $ is given by
\be
H_{\kappa}^\xi = \frac{p_{\xi}^2}{2}
+\frac{\omega^2}{2\kappa\gamma^2\Ck_{\kappa}^2(\xi)} , \qquad   \qquad \{ H_{\kappa} , H_{\kappa}^\xi \}=0.
\label{hc31}
\ee

Therefore, $H_{\kappa}$ defines an integrable system for any value of $\omega$ and $\gamma$.  We remark that the integral   $H_{\kappa}^\xi$ is, in fact, a one-dimensional Higgs-type oscillator (\ref{da}) with ``frequency'' $\omega/\gamma$ on the variable $\xi$,
since it can be rewritten through (\ref{prop})  as
\be
H_{\kappa}^\xi= \frac{p_{\xi}^2}{2}
+\frac{\omega^2}{2 \gamma^2}{\Tk_{\kappa}^2(\xi)}+\frac{\omega^2}{2 \k\gamma^2}.
\nonumber
\ee
Note that its Euclidean limit $H^\xi$ (\ref{af}) is recovered in the form
$$
\lim_{\k\to 0} \left( H_{\kappa}^\xi -\frac{\omega^2}{2 \k\gamma^2} \right) = \frac12\, p_\xi^2 + \frac{\omega^2}{2\gamma^2} \xi^2  .
$$
 
In what follows, we will factorize  the classical Hamiltonians $H_{\kappa}^\xi$ and  $H_{\kappa}$ for a generic value of $\gamma$, by taking into account that expressions
  (\ref{hc3})  and  (\ref{hc31})  correspond to classical P\"oschl--Teller
Hamiltonians, whose factorization properties have been previously considered in~\cite{Kuru1}.
Moreover, when $\gamma$ is a rational number the factorization approach will lead us to the superintegrability of the complete system 
(\ref{hc3}).

 %%%%%%%%%%%%%%%%%%%%%%%%%%%%%%%%%%%%%%%%%%%%%%%%%%%%%%55
\subsect{Ladder functions}

Firstly, we search for some functions  $B_\k^\pm(\xi)$ that generate a Poisson algebra of the type
\be
\{H_{\kappa}^\xi , B_\k^\pm\}=\mp i\,f(H_{\kappa}^\xi) B_\k^\pm ,
\nonumber
\ee
for some function $f$. They will be the so-called  {\em ladder} functions for $H_{\kappa}^\xi$, and can be found to be~\cite{Kuru1}
\begin{equation}\label{bc}
B_\k^{\pm} ={\mp}\frac{i}{\sqrt{2}}\Ck_{\kappa}(\xi)\, p_{\xi}+
\frac{\mm}{\sqrt{2}} \Sk_{\kappa}(\xi),
\end{equation}
where 
\be
\mm(p_{\xi},\xi)= \sqrt{2\kappa H_{\kappa}^\xi} \, .
\label{newfunction}
\ee
In fact, it is straightforward to check that
\be\label{factbp}
H_{\kappa}^\xi=B_\k^+B_\k^- + \frac{\omega^2}{2\kappa\gamma^2},
\ee
and the following Poisson algebra is obtained
\begin{equation}\label{commpt1}
\{H_{\kappa}^\xi,B_\k^{\pm}\}= \mp i\,\mm\,B_\k^{\pm} ,\qquad \{ B_\k^-,B_\k^+\}=- i  \,\mm \, .
\end{equation}
As a  consequence,
\begin{equation}\label{commpt2}
\{H_{\kappa} ,B_\k^{\pm}\}=\mp  i\,\frac{\gamma^2 \mm}{\Ck_{\kappa}^2(y)}\,B_\k^{\pm},
\end{equation}
where  from~\eqref{hc31} we know that the function $\mm$  (\ref{newfunction}) is a constant of the motion for $H_{\kappa}$.

 %%%%%%%%%%%%%%%%%%%%%%%%%%%%%%%%%%%%%%%%%%%%%%%%%%%%%%55
\subsect{Shift functions}

Next, we look for {\em shift} functions $A_\k^\pm(y)$ that factorize $H_{\kappa}$. This implies the search for a Poisson algebra of the type
\be
\{H_{\kappa} , A_\k^\pm\}
=\pm i\,g(\mm,y) A_\k^\pm ,
\nonumber
\ee
for a certain function $g$ including the potential in (\ref{hc3}). By taking into account that $\mm$  (\ref{newfunction}) is a constant of the motion for $H_{\kappa} $  (\ref{hc3}) and by imposing  that the Hamiltonian can be factorized in the form
\begin{equation}\label{cht2}
H_{\kappa} =A_\k^{+} A_\k^{-}+\lambda^A_{\k} ,
\end{equation}
we obtain the shift functions
\begin{equation}\label{capm}
A_\k^{\pm}=\mp \frac{i}{\sqrt{2}}\,p_{y}-\frac{\gamma \mm}{\sqrt{2} }\Tk_{\kappa}(y),\qquad 
\lambda^A_{\k} =\frac{1}{2\kappa}  \left(\gamma^2\mm^2-\omega^2\right) ,
\end{equation}
and the Poisson algebra
\begin{equation}\label{commpt3}
\{H _{\kappa},A_\k^{\pm}\}=\pm i\,\frac{\gamma\mm}{\Ck_{\kappa}^2(y)}\,A_\k^{\pm} ,\qquad \{ A_\k^-,A_\k^+\}=i\,\frac{\gamma\mm}{\Ck^2_{\kappa}(y)} .
\end{equation}

 %%%%%%%%%%%%%%%%%%%%%%%%%%%%%%%%%%%%%%%%%%%%%%%%%%%%%%55
\subsect{Additional symmetries}

  The superintegrability of the Hamiltonian $H _{\kappa}$ for rational values $\gamma=m/n$ is now easily deduced from the factorization approach. In fact, 
  from (\ref{commpt2}) and (\ref{commpt3}), a straightforward computation shows that the ladder and shift functions provide two additional integrals of  the motion    for   $H_{\kappa} $:
\be
\{ H_{\kappa} , X_\k^{\pm}\}=0, \qquad \mbox{where} \qquad X_\k^{\pm}=(B_\k^{\pm})^n (A_\k^{\pm})^{m}\,. 
\label{csymmet1}
\ee
From the factorization properties of $ B^\pm$ and $ A^\pm$ given in
(\ref{factbp}) and (\ref{cht2}), we conclude that 
\begin{equation}\nonumber
X^+ X^- = \left( H_\k^\xi - \frac{\omega^2}{2\k \gamma^2} \right)^n 
\left(H_\k -\gamma^2 \left(  H_\k^\xi-\frac{\omega^2}{2\k \gamma^2}\right) \right)^m \,.
\end{equation}
Therefore, the four   constants of motion $(H_\k,H_\k^\xi,X_\k^\pm)$ are functionally
dependent. However, any of the sets  $(H_\k,H_\k^\xi,X_\k^+)$
or $(H_\k,H_\k^\xi,X_\k^-)$ is formed by functionally independent  integrals.

In principle the constants of motion $X_\k^\pm$ are complex and they include
powers of the square root of $\mm$ (\ref{newfunction}). We have two situations giving rise
to real constants of motion $\xx_\k$ and $\yy_\k$:
\begin{itemize}
\item[i)] If $m+n$ is even we have
\begin{equation}\label{even}
X_\k^\pm = \pm i \, \mm \yy_\k  + \xx_\k .
\end{equation}
\item[ii)] If $m+n$ is odd we find
\begin{equation}\label{odd}
X_\k^\pm = \mm\xx_\k  \pm i \,\yy_\k .
\end{equation}
\end{itemize}
The symmetries $\xx_\k$ and $\yy_\k$ are polynomial in the momenta, whose degrees are $m+n$ and $m+n-1$ for case i), and $m+n-1$ and $m+n$ in case ii), respectively
\cite{lissajous}. 
It can also be proven that the algebraic structure generated by the Poisson brackets 
of the sets of integrals of motion $(H_\k,H_\k^\xi,X_\k^\pm)$ or
$(H_\k,H_\k^\xi,\xx_\k,\yy_\k)$  also  gives 
rise to a polynomial algebra, as in the Euclidean case. Therefore, the generalization of  Theorem 1 to the sphere and the hyperbolic space can be stated as follows:

\medskip

\noindent
{\bf Theorem 3.} {\em {\rm (i)} For any value of the real anisotropy parameter $\gamma$, the Hamiltonian $H_\k$ (\ref{hc1}) defines an   integrable anisotropic curved oscillator on ${\mathbf S}^2$   and $ {\mathbf H}^2$, whose  (quadratic) constant of motion is given by   $H^\xi_\k$   (\ref{hc31}).

\noindent
 {\rm (ii)} When $\gamma=m/n$ is a rational parameter,    $H_\k$  defines   a  superintegrable  anisotropic curved oscillator  and the additional constant of motion  is given by either $\xx_\k$ or $\yy_\k$ in (\ref{even}) and (\ref{odd}). The   sets $(H_\k,H_\k^\xi,\xx_\k)$ and $(H_\k,H_\k^\xi,\yy_\k)$ are formed by three functionally independent functions.
}

As far as the  (flat) Euclidean  limit $\k\to 0$ is concerned we remark that,  despite the expressions (\ref{hc3}) and  (\ref{hc31})    are only defined if $\k\ne 0$, all the remaining ones have a well defined Euclidean limit.  The latter can be achieved by taking into a account the following limits  of the integrals $H_{\kappa}^\xi$ (\ref{hc31}) and $\mm$ (\ref{newfunction})   
\be
\lim_{\k\to 0} \k H_{\kappa}^\xi=\frac{\omega^2}{2\gamma^2},\qquad \lim_{\k\to 0} \mm=\frac{\omega}{\gamma} .
\label{sb}
\ee
Hence, when $\k=0$,  we find that the curved  Hamiltonian $H_{\kappa}$ (\ref{hc1})  reduce to $H$  (\ref{ae}), the curved ladder functions $B_{\kappa}^\pm$    (\ref{bc})   to $B^\pm$    (\ref{bc0}), the curved shift functions $A_{\kappa}^\pm$    (\ref{capm})   to $A^\pm$    (\ref{capm0}),  $\lambda^A_{\k}$   to $\gamma^2 H^\xi$ (\ref{af}), and the curved integrals $X_\k^{\pm}$ (\ref{csymmet1}) to $X^{\pm}$ (\ref{csymmet10}).

 %%%%%%%%%%%%%%%%%%%%%%%%%
\subsect{Examples}

We now  illustrate the results described by Theorem 3 through the   particular cases with $\gamma=\{1,2,1/2\}$. The first two ones   generalize the Euclidean oscillators presented  in Sections 2.1 and 2.2, while the third one allows us to show explicitly the non-equivalence among the curved potentials $U_\k^\gamma$ and  $U_\k^{1/\gamma}$, despite of the fact that when $\k=0$ both of them lead to equivalent Euclidean potentials.
 
 %%%%%%%%%%%%%%%%%%%%%%%%%
 \subsubsect{The  $\gamma=1$ curved (Higgs) oscillator}

We set   $\gamma=m=n=1$ so that  $\xi=x$ and $p_\xi=p_x $. The Hamiltonian  $H_\k$ (\ref{hc1}) (with potential $U_\k^{\gamma=1}=U_\k^{1:1}$ (\ref{db})) and also in the form (\ref{hc3}) reads
\be
{H}_{\kappa}^{\gamma=1} =   \frac{1}{2} \left(\frac{p_{x}^2}{ \Ck_\k^2(y)}+p_{y}^2\right)+\frac{\omega^2}{2} 
\left(\frac{ \Tk_{\kappa}^2(  x)}{ \Ck_{\kappa}^2(y)}+\Tk_{\kappa}^2(y)\right)  = \frac{p_{y}^2}{2}+\frac{  H_{\kappa}^x}{\Ck_{\kappa}^2(y)}-
\frac{\omega^2}{2\kappa}  ,
\nonumber
\ee
where the quadratic  integral $H_{\kappa}^x\equiv H_{\kappa}^\xi$ is given by 
\be
H_{\kappa}^x= \frac{p_{x}^2}{2}
+\frac{\omega^2 }{2\kappa\Ck_{\kappa}^2(x)} .
\nonumber
\ee
The polynomial integrals (\ref{even}) turn out to be
\bea
&& \xx_\k=-\frac 12\left(  \Ck_{\kappa} (x) p_xp_y+\mm^2 \Sk_{\kappa} (x)\Tk_{\kappa} (y)\right) ,\nonumber\\[2pt]
&&\yy_\k=-\frac 12\,\left(  \Sk_{\kappa} (x) p_y- \Ck_{\kappa} (x)\Tk_{\kappa} (y) p_x\right) .
\label{sa}
\eea

Notice that the integral $\yy_\k$ is proportional to the (curved) angular momentum $\ang_\k$ which in geodesic parallel and polar variables can be shown to be given by~\cite{mariano99,conf}
\be
\ang_\k =\Sk_{\kappa} (x) p_y- \Ck_{\kappa} (x)\Tk_{\kappa} (y) p_x= p_\phi .
\label{allb}
\ee

Under the flat limit $\k\to 0$, we recover the results of the Euclidean isotropic oscillator given in Section 2.1. In particular, the integrals $\xx_\k$
and $\mm \yy_\k$ from (\ref{sa}) and the curved angular momentum (\ref{allb}) reduce to   (\ref{al}) and  (\ref{all}), since $\mm \to \omega$.

 %%%%%%%%%%%%%%%%%%%%%%%%%
  \subsubsect{The  $\gamma=2$  curved  oscillator}

In this case, we choose $\gamma=m=2$ and $n=1$ so that $\xi=2x$ and $p_\xi =p_x/2 $. Thus the Hamiltonian $H_\k$ (\ref{hc1}) with potential $U_\k^{\gamma=2}=U_\k^{2:1}$ (\ref{dd}) is given by
\be
{H}_{\kappa}^{\gamma=2}  =   \frac{1}{2} \left(\frac{p_{x}^2}{ \Ck_\k^2(y)}+p_{y}^2\right)+\frac{\omega^2}{2} 
\left(\frac{ \Tk_{\kappa}^2( 2 x)}{ \Ck_{\kappa}^2(y)}+\Tk_{\kappa}^2(y)\right)  = \frac{p_{y}^2}{2}+\frac{ 4 H_{\kappa}^\xi}{\Ck_{\kappa}^2(y)}-
\frac{\omega^2}{2\kappa}  ,
\label{pb}
\ee
where 
\be
H_{\kappa}^\xi= \frac{p_{\xi}^2}{2}
+\frac{\omega^2 }{8\kappa\Ck_{\kappa}^2(\xi)}= \frac{p_{x}^2}{8}
+\frac{\omega^2 }{8\kappa\Ck_{\kappa}^2(2x)} .
\nonumber
\ee
From the additional symmetries $X_\k^\pm$   (\ref{csymmet1}) we find that the
 polynomial integrals (\ref{odd}) read
\bea
&& \xx_\k=
-\frac {1}{2\sqrt{2}}\left( \left[   \Sk_{\kappa} (2x)p_y-2  \Ck_{\kappa} (2x)  \Tk_{\kappa} (y) p_x  \right] p_y     - 4 \mm^2   \Sk_{\kappa} (2x)   \Tk^2_{\kappa} (y)\right)  ,
\label{pol21aclas}\\[2pt]
&&\yy_\k=
\frac {1}{4\sqrt{2}}\left(      
        \Ck_{\kappa} (2x)    p_x p_y^2 + 4 \mm^2\Tk_{\kappa} (y) \left[ 2\Sk_{\kappa} (2 x)p_y -  \Ck_{\kappa} (2 x)     \Tk_{\kappa} (y)  p_x   \right]          \right).
       \label{pol21bclas}
\eea

 If $\k\to 0$, the limit   (\ref{sb})  yields $\mm \to \omega/2$ and the  integrals $\mm \xx_\k$ and $\yy_\k$ reduce to (\ref{am}), thus reproducing the results of Section 2.2.

 %%%%%%%%%%%%%%%%%%%%%%%%%
   \subsubsect{The  $\gamma=1/2$  curved  oscillator}

 Now we fix  $\gamma=1/2$,  $m=1$ and $n=2$. Then $\xi= x/2$, $p_\xi =2 p_x $ and  the corresponding Hamiltonian $H_\k$ (\ref{hc1}) with potential $U_\k^{\gamma=1/2}$ (\ref{de}) is given by
\be
{H}_{\kappa}^{\gamma=1/2}  =   \frac{1}{2} \left(\frac{p_{x}^2}{ \Ck_\k^2(y)}+p_{y}^2\right)+\frac{\omega^2}{2} 
\left(\frac{ \Tk_{\kappa}^2( \frac x2)}{ \Ck_{\kappa}^2(y)}+\Tk_{\kappa}^2(y)\right)  = \frac{p_{y}^2}{2}+\frac{  H_{\kappa}^\xi}{4\Ck_{\kappa}^2(y)}-
\frac{\omega^2}{2\kappa}  ,
\label{pa}
\ee
where 
\be
H_{\kappa}^\xi= \frac{p_{\xi}^2}{2}
+\frac{2\omega^2 }{ \kappa\Ck_{\kappa}^2(\xi)}= 2  {p_{x}^2} 
+\frac{2\omega^2 }{ \kappa\Ck_{\kappa}^2(\frac x2)} .
\nonumber
\ee
Notice that, due to the term $\Tk_{\kappa}^2( \frac x2)$ in the potential, the Hamiltonian (\ref{pa}) defines a different/non-equivalent curved oscillator to  the  previous $\gamma=2$ case (\ref{pb}) which involved the term $\Tk_{\kappa}^2( 2x)$ in the potential.

The  additional integrals (\ref{odd})  for ${H}_{\kappa}^{\gamma=1/2} $ read
\bea
&& \xx_\k=
-\frac {1}{4\sqrt{2}}\left( 4\left[   \Sk_{\kappa} (x)p_y-   \Ck^2_{\kappa} (\tfrac x2)  \Tk_{\kappa} (y) p_x  \right] p_x     + \mm^2   \Sk^2_{\kappa} (\tfrac x2)   \Tk_{\kappa} (y)\right)  ,\label{clas12a}\\[2pt]
&&\yy_\k=
\frac {1}{2\sqrt{2}}\left(      
      4  \Ck^2_{\kappa} (\tfrac x2)    p^2_x p_y - \mm^2  \left[  \Sk^2_{\kappa} (\tfrac x2)p_y -  \Sk_{\kappa} (  x)     \Tk_{\kappa} (y)  p_x   \right]          \right).
      \label{clas12b}
\eea

We finally remark that the above three  anisotropic curved oscillators are the only ones within the family (\ref{hc1}) which are quadratic (in the momenta) superintegrable systems. All the remaining ones with a rational $\gamma$, are also superintegrable Hamiltonians but the additional integral is always of higher-order in the momenta. Some trajectories on the sphere and on the hyperboloid for these three systems are plotted in Fig.~2.

%%%%%%%%%%%%%% figure 2%%%%%%%%%%%%%%%%%%%%%%%%%%

\begin{figure}[t]
\setlength{\unitlength}{1mm}
\begin{picture}(135,120)(0,0)
\footnotesize{
\put(10,65){\includegraphics[scale=0.35]{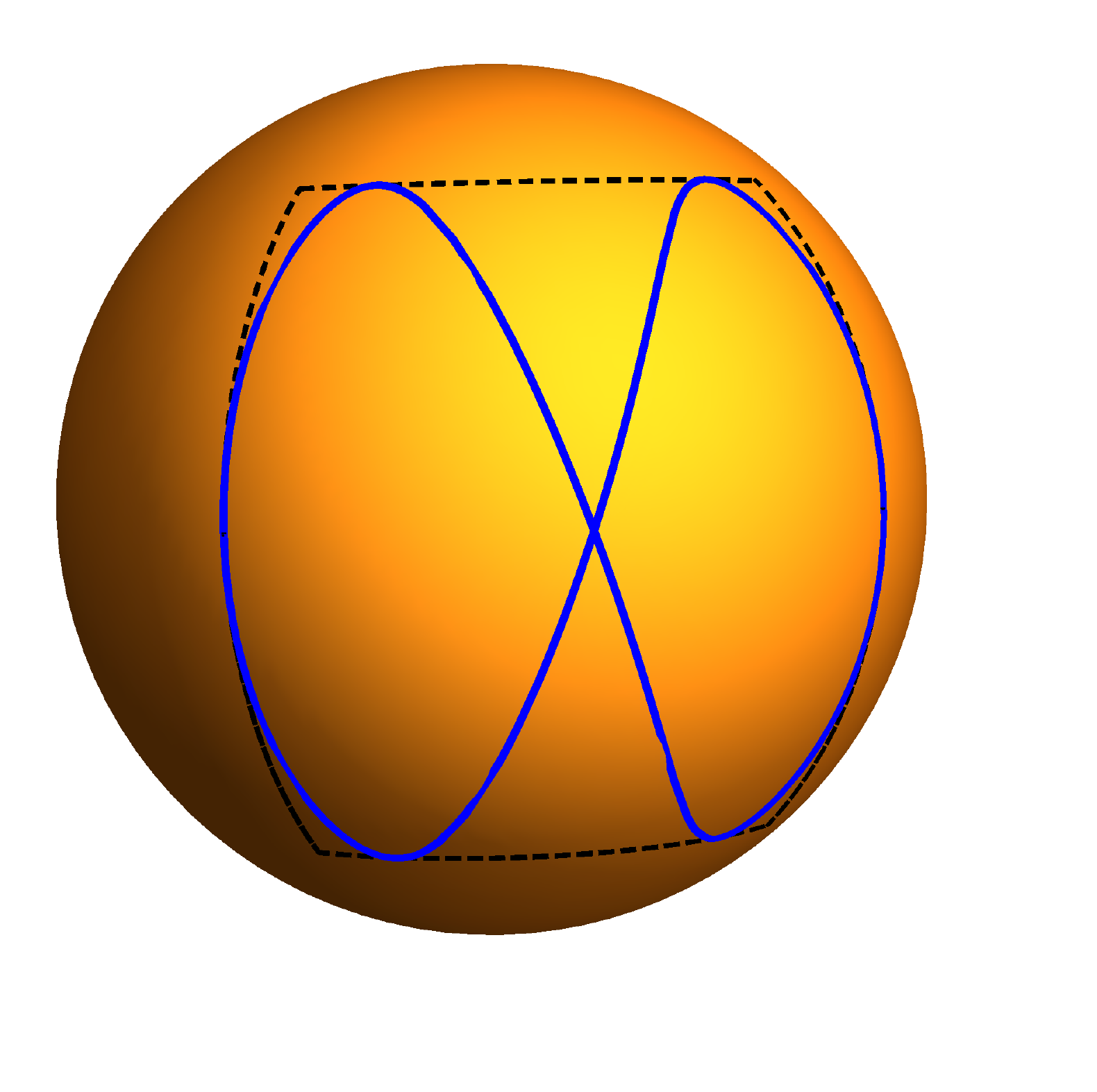}}
\put(34,64){(a)}
\put(90,64){\includegraphics[scale=0.35]{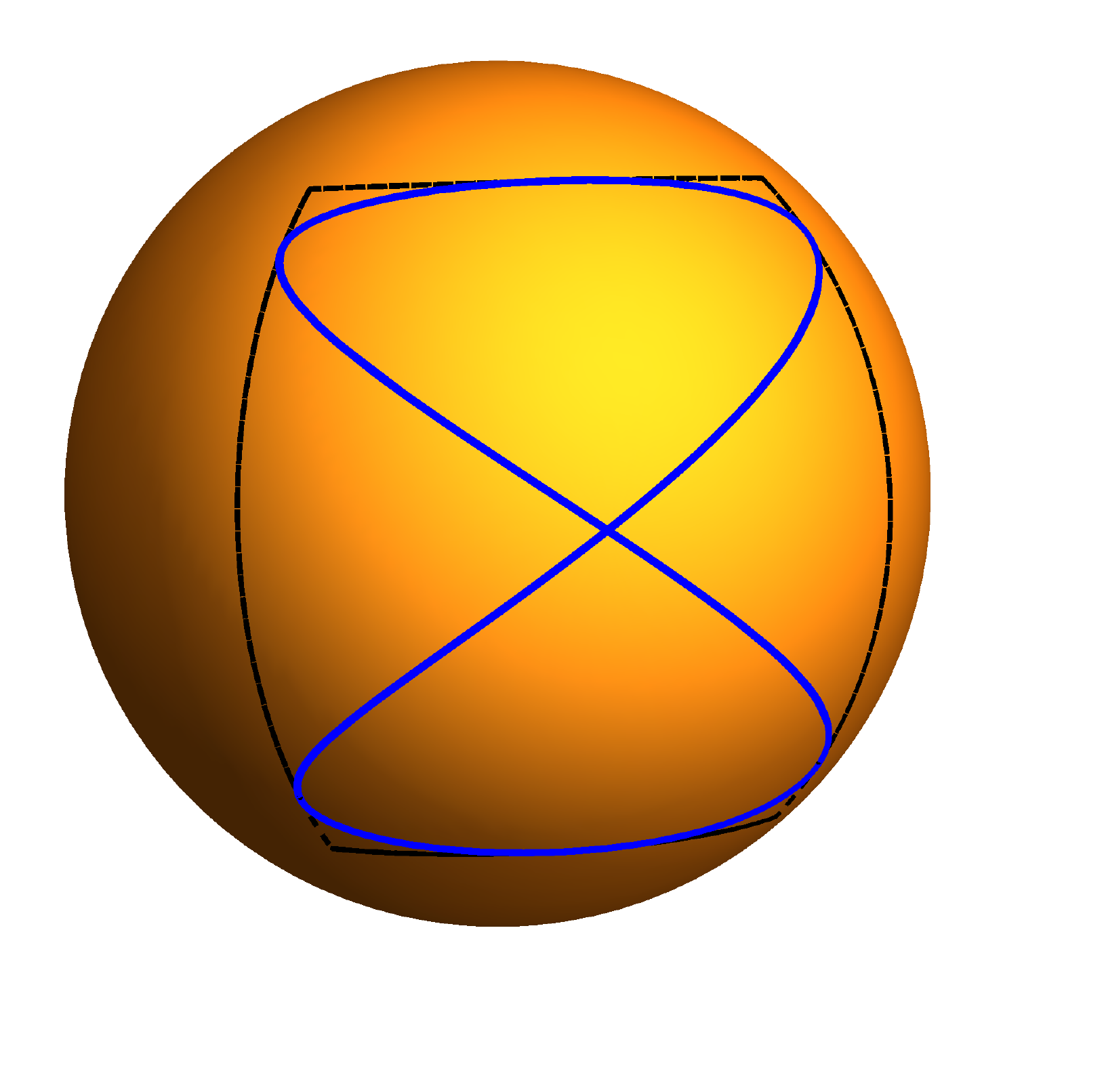}}
\put(115,64){(b)}
\put(7,10){\includegraphics[scale=0.28]{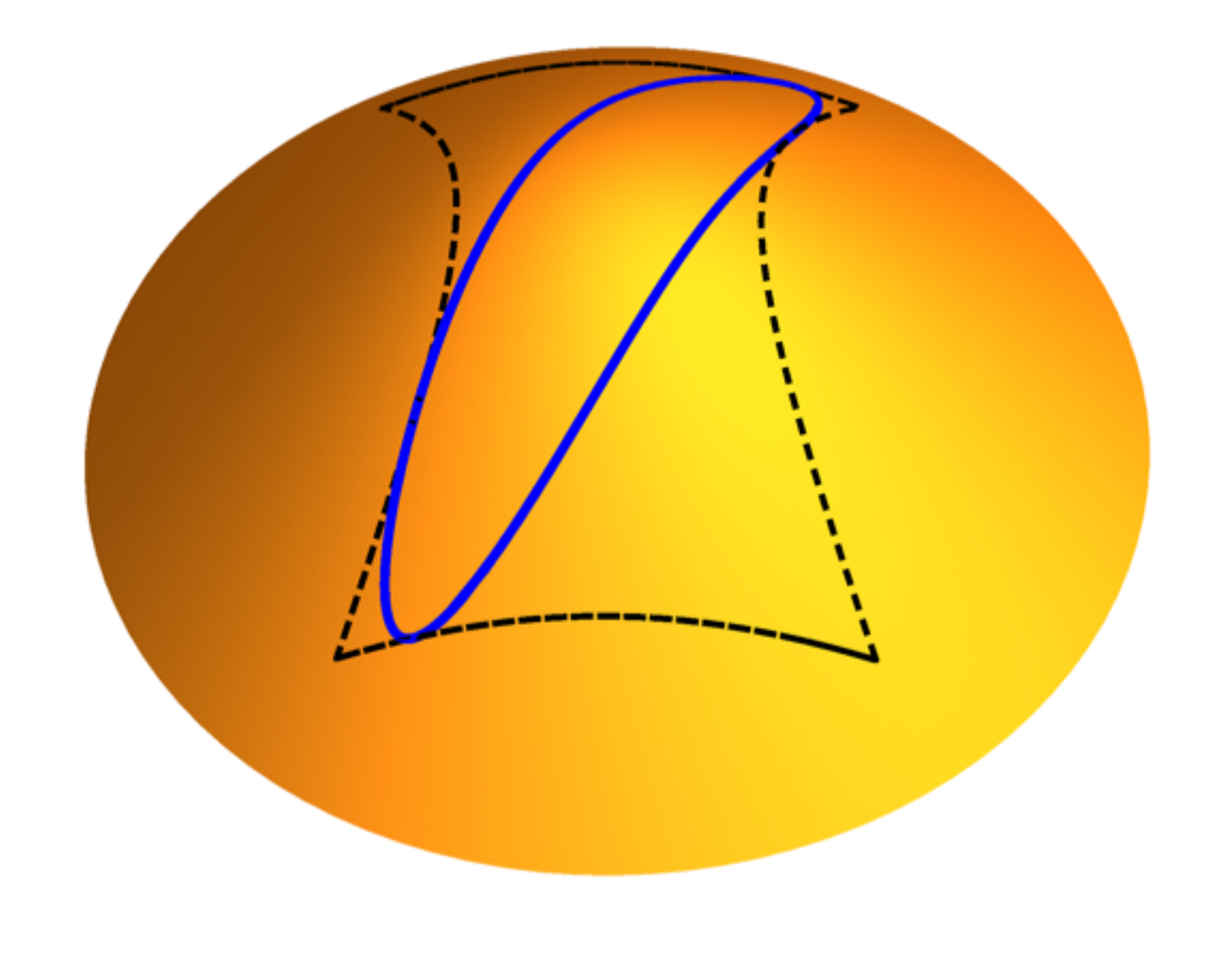}}
\put(34,6){(c)}
\put(85,13){\includegraphics[scale=0.4]{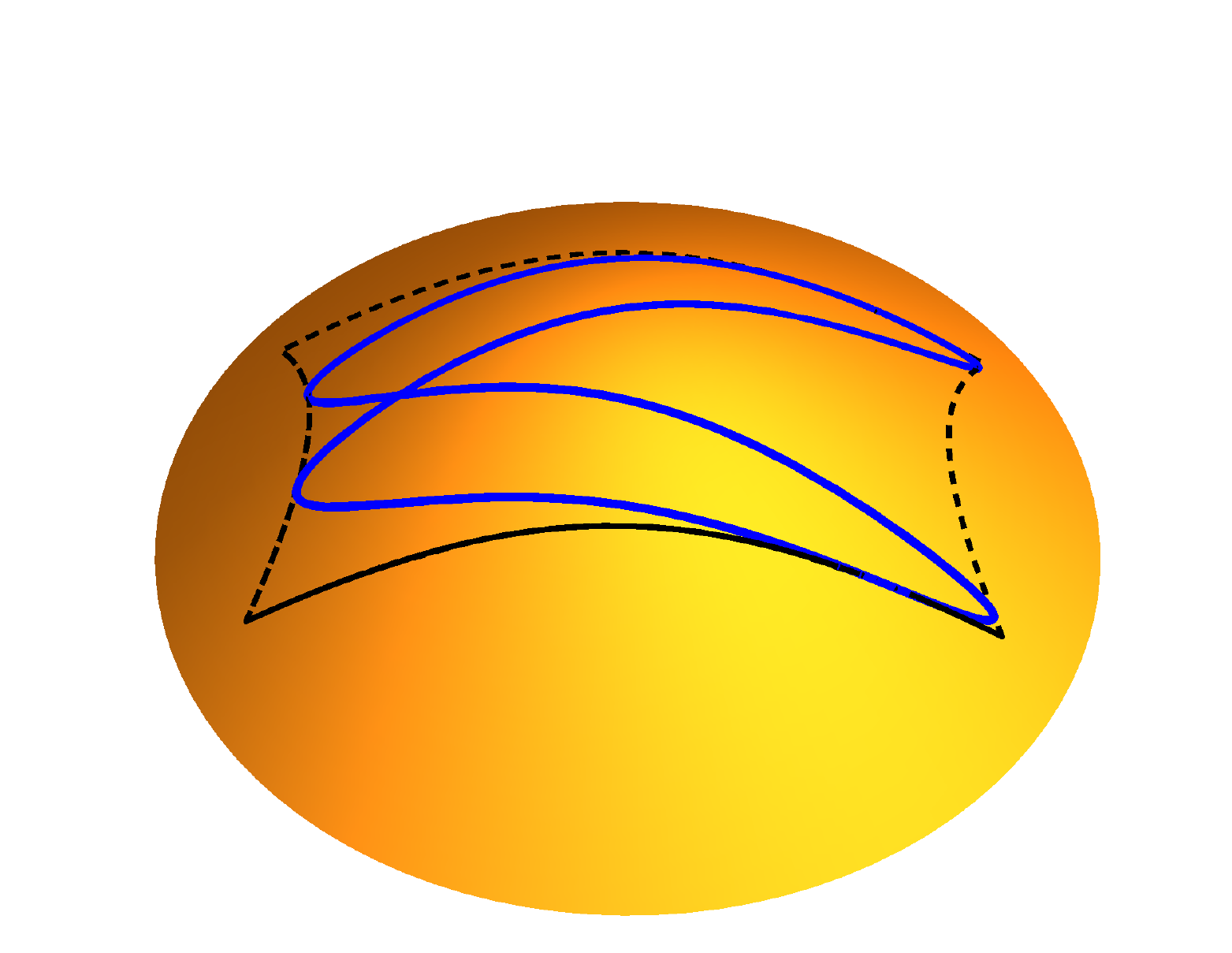}}
\put(115,6){(d)}
}
\end{picture}
\caption{\footnotesize Examples of trajectories for the Hamiltonian $H_\k$ (\ref{hc1}): (a) on the sphere for $\gamma=1/2$, (b) 
on the sphere for $\gamma=2$,  (c) on the hyperboloid  for $\gamma=1$, and (d) 
 on the hyperboloid  for $\gamma=2$. }  
 \end{figure}
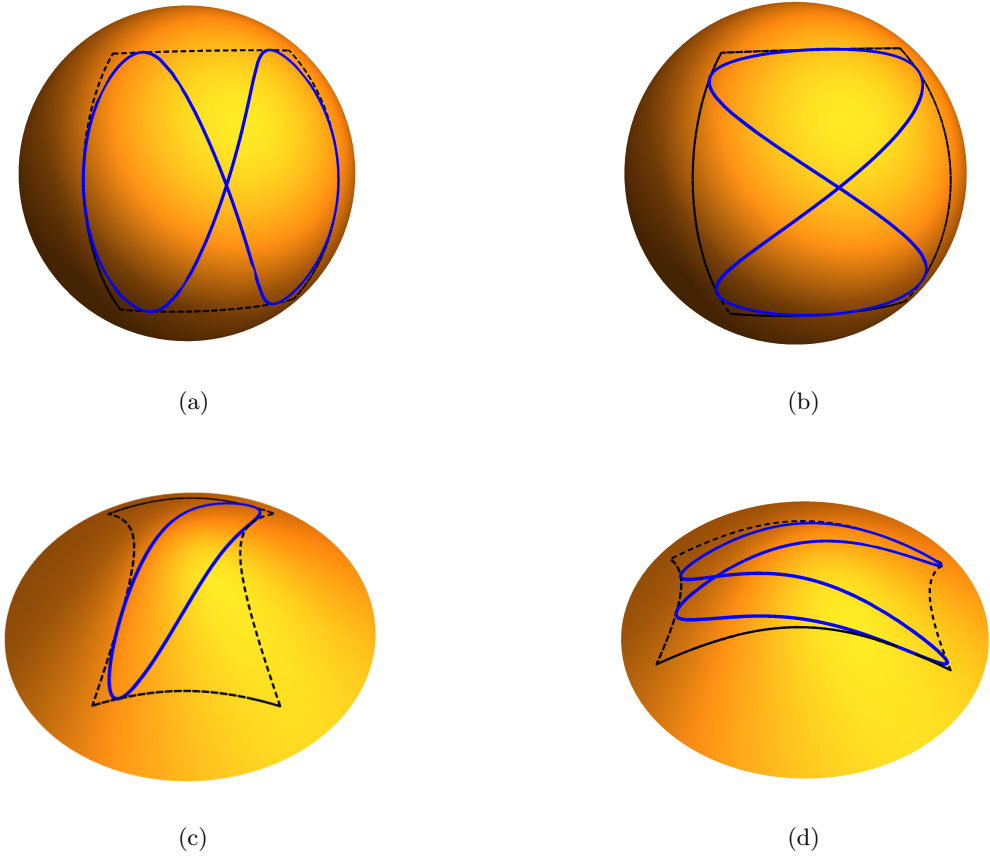
\ 
%%%%%%%%%%%%%%%%%%%%%%%%%%%%%%%%%%%%%%%%%%%%%%%%%%%%%%

%%%%%%%%%%%%%%%%%%%%%%
\sect{Quantum   anisotropic curved oscillators}

In order to construct the quantum analogue of the   Hamiltonian $H_\k$ (\ref{hc1}), let us   consider the 
 Laplace--Beltrami  (LB) operator on a two-dimensional (curved) space
 \be
\Delta_{\rm LB}=\sum_{i,j=1}^2 \frac 1{\sqrt{g}}\partial_i\sqrt{g} g^{ij}\partial_j ,
\label{LBop}\nonumber
\ee
where $g^{ij}$ is the inverse of the metric tensor $g_{ij}$ and $g$ is the   determinant. In terms of the geodesic parallel and geodesic polar coordinates 
(see Eq.~(\ref{fa}) in the Appendix A) we obtain the following Laplacian operator on   ${\mathbf S}^2$   and $ {\mathbf H}^2$ with curvature parameter $\k$~\cite{conf}:
\be
\Delta_{\rm LB}=\frac{1}{\Ck_{\kappa}^2(y)} \,\frac{\partial^2}{\partial x^2}+
\frac{\partial^2}{\partial y^2} - \kappa \Tk_{\kappa}(y)\frac{ \partial}{\partial y} =  \frac{\partial^2}{\partial r^2}+     
\frac{1}{\Tk_{\kappa}(r)} \,\frac{\partial}{\partial r}+   \frac{1}{\Sk^2_{\kappa}(r)} \,\frac{\partial^2}{\partial \phi^2}  .
\nonumber
\ee
Next, if we assume the so-called LB quantization prescription on curved spaces (see~\cite{t1, darboux, taub, Chanu} and references therein) the free quantum  Hamiltonian of a unit mass particle will read
\be
\hat{\cal T}_\k=-\frac {\hbar^2} 2 \Delta_{\rm LB} .
\nonumber
\ee
Now, we can define the  quantum curved anisotropic oscillator Hamiltonian as $\hat{H}_\k=\hat{\cal T}_\k+\hat{U}^\gamma_\k$  with the potential (\ref{de}), 
which in terms of geodesic parallel coordinates is given by
\begin{equation}\label{hc11}
\hat  H_{\kappa} = -\frac{\hbar^2}{2} 
\left(\frac{1}{\Ck_{\kappa}^2(y)} \,\frac{\partial^2}{\partial x^2}+
\frac{\partial^2}{\partial y^2} - \kappa \Tk_{\kappa}(y)\frac{ \partial}{\partial y}\right)+
\frac{\omega^2}{2} 
\left(\frac{ \Tk_{\kappa}^2(\gamma x)}{ \Ck_{\kappa}^2(y)}+\Tk_{\kappa}^2(y)\right),
\end{equation}
where  $\gamma$ and $\omega$ are real positive parameters and the domains of
$x$ and $y$ are shown in (\ref{range}). 
 
By applying the relation (\ref{prop}) to the terms $\Tk_{\kappa}^2(\gamma x)$ and $\Tk_{\kappa}^2(y)$ and
after the change of variable $\xi={\gamma x}$ with domain \eqref{rangeS} and~\eqref{range}, the quantum Hamiltonian reads 
\begin{equation}\label{hq3}
\hat{H}_{\kappa} = -\frac{\hbar^2}{2}\,\frac{ \partial^2}{\partial y^2}+\frac{\hbar^2}{2} \kappa \Tk_{\kappa}(y)\frac{ \partial}{\partial y}
+\frac{\gamma^2}{\Ck_{\kappa}^2(y)}\left(-\frac{\hbar^2}{2}\frac{\partial^2}{\partial \xi^2}+\frac{\omega^2}{2 \kappa \gamma^2 \Ck_{\kappa}^2(\xi)}\right)-
\frac{\omega^2}{2 \kappa} ,\quad\ \k\ne 0 .
\end{equation} 
Thus, we can write $\hat{H}_{\kappa} $ in  terms of a one-dimensional symmetry operator $\hat{H}_{\kappa}^{\xi}$ such that $[\hat{H}_{\kappa},\hat{H}_{\kappa}^{\xi}]=0 $, namely
\bea
&&\hat{H}_{\kappa}= -\frac{\hbar^2}{2}\,\frac{ \partial^2}{\partial y^2}+\frac{\hbar^2}{2} \kappa \Tk_{\kappa}(y)\frac{ \partial}{\partial y}
+\frac{\gamma^2 \hat{H}_{\kappa}^{\xi}}{\Ck_{\kappa}^2(y)}-
\frac{\omega^2}{2 \kappa}  ,\qquad\k\ne 0  , \label{hq1}\\[2pt]
&&\hat{H}_{\kappa}^{\xi}= -\frac{\hbar^2}{2}\,\frac{\partial^2}{\partial \xi^2}+\frac{\omega^2}{2 \kappa \gamma^2 \Ck_{\kappa}^2(\xi)}.
\label{hq} 
\eea
Remark that  $\hat H_{\kappa}^\xi$, which is the quantization of~\eqref{hc31},  is just the quantum P\"oschl--Teller Hamiltonian written simultaneously in its trigonometric $(\kappa>0)$ and hyperbolic $(\kappa < 0)$ versions (see~\cite{PT} and references therein). Note that 
the Euclidean oscillator in $\xi$ is obtained under the limit
$$
\lim_{\k\to 0} \left( \hat H_{\kappa}^\xi -\frac{\omega^2}{2 \k\gamma^2} \right) =  -\frac{\hbar^2}{2}\frac{\partial^2}{\partial \xi^2} + \frac{\omega^2}{2\gamma^2} \,\xi^2  .
$$

Now, the eigenvalue equation for $\hat{H}_{\kappa}$ that we want to solve  is  
\begin{equation}\label{eve1}
\hat{H}_{\kappa}\Psi_\k(\xi,y)=E_{\kappa}\Psi_\k(\xi,y) ,
\end{equation}
where we are looking for factorizable solutions in the form
$\Psi_\k(\xi,y)=\Xi_\k^\mmb(\xi)\,Y_\k^{\gamma\mmb}(y)$. If the function $\Xi_\k^\mmb(\xi)$ fulfills the eigenvalue equation
\be
\hat{H}_{\kappa}^{\xi} \,\Xi_\k^\mmb(\xi)=E_{\kappa}^{\xi}\,\Xi_\k^\mmb(\xi),
\qquad
\mbox{with}
\qquad
\mmb=\sqrt{2\kappa E_{\kappa}^{\xi}},
\label{eve3}
\ee
then the second component $Y_\k^{\gamma\mmb}(y)$ of the solutions for~\eqref{eve1} can be obtained as a one-dimensional eigenvalue problem
\be
\hat{H}_{\kappa}Y_\k^{\gamma\mmb}(y)=E_{\kappa} Y_\k^{\gamma\mmb}(y)
\label{eve4}
\ee
 for the Hamiltonian
\be
\hat{H}_{\kappa}= -\frac{\hbar^2}{2}\,\frac{ \partial^2}{\partial y^2}+\frac{\hbar^2}{2} \kappa \Tk_{\kappa}(y)\frac{ \partial}{\partial y}
+\frac{(\gamma  \mmb)^2}{2\k \Ck_{\kappa}^2(y)}-
\frac{\omega^2}{2 \kappa}  ,\qquad\k\ne 0  .
\label{Hk}
\ee

In particular, we shall deal separately with each of the two
one-dimensional problems (\ref{eve3}) and (\ref{eve4}) by means of  the factorization approach \cite{infeldhull51, Kuru1}.
In this way we will find   ladder operators  $\hat B^\pm_\k$
for $\hat{H}_{\kappa}^{\xi}$ (\ref{hq}) and   shift operators  $\hat A^\pm_\k$ for
$\hat{H}_{\kappa}$  (\ref{hq1}) or (\ref{Hk}).  As in the classical case,  we will be able to   deduce  ``additional" quantum symmetries  for $\hat H_\kappa$ (\ref{hq3}) 
when $\gamma$ be a rational number.

%%%%%%%%%%%%%%%%%%%%%%%%%%%%%%%%%%%%
\subsect{Ladder operators for the Hamiltonian $\hat{H}_{\k}^{\xi}$}

In order to find the ladder operators $\hat B^\pm_{\k}$ for  $\hat{H}_{\k}^{\xi}$ (\ref{hq}), 
 we express the  
corresponding  eigenvalue equation (\ref{eve3})  as
\begin{equation}
\left(         -\frac{\hbar^2}{2}\Ck_{\kappa}^2(\xi)\,\frac{\partial^2}{\partial \xi^2}  -\Ck_{\kappa}^2(\xi)\,E_{\kappa}^{\xi}\right) \Xi_\k^\mmb(\xi)=\left(-\frac{\omega^2}{2\kappa\gamma^2} \right) \Xi_\k^\mmb(\xi) \,.
\nonumber
\end{equation}
Now, we define the diagonal operator $\hat\mm$ that acts on the space of eigenfunctions $\Xi_\k^\mmb(\xi)$ in the form
\be\label{e}
\hat\mm\, \Xi_\k^\mmb(\xi)=\mmb\, \Xi_\k^\mmb(\xi)\,
\ee 
where $\mmb$ was defined in (\ref{eve3}). 
Next we define the operator   
\begin{equation}\label{hqhh}\nonumber
\hat{h}_{\k} =   -\frac{\hbar^2}{2}  {\Ck_{\kappa}^2(\xi)} \frac{ \partial^2}{\partial\,\xi^2}-\Ck_{\kappa}^2(\xi)\,\frac{(\hat\mm)^2}{2\kappa} ,
\end{equation}
that can be factorized in terms of two first-order $\hat\mm$--dependent differential operators plus another diagonal operator in the form
\begin{equation}\label{factb}
\hat{h}_{\k}=\hat{B}_{\k}^-\hat{B}_{\k}^+ +\hat\lambda^B_{\k} \, ,
\end{equation}
where
\begin{equation}
\begin{array}{l}
\displaystyle \hat{B}_{\k}^- =\frac{\hbar}{\sqrt{2}} {\Ck_{\kappa}(\xi)} \frac{\partial}{\partial\,\xi}+\frac 1{\sqrt{2}}  \Sk_{\kappa}(\xi)
\,\hat\mm   ,
\\[2.ex]
\displaystyle \hat{B}_{\k}^+=-\frac{\hbar}{\sqrt{2}} {\Ck_{\kappa}(\xi)} \frac{\partial}{\partial\,\xi}+\frac 1{\sqrt{2}}   \Sk_{\kappa}(\xi)\, \hat\mm,
\\[2.ex]
\displaystyle \hat\lambda^B_{\k}=-\frac{\hat\mm}{2\kappa}(\hat\mm +\hbar{\kappa} ) .
\label{lep}\nonumber
\end{array}
\ee

These  operators   can be called {\em pure-ladder} ones in order to
stress that they correspond to different eigenvalues for $\hat{H}_{\kappa}^{\xi}$ (see~\cite{kuru09}) and their action on the eigenfunctions of  $\hat{H}_{\kappa}^{\xi}$ is straightforwardly shown to be 
\begin{equation}\nonumber
\hat{B}_\k^+ \Xi_{\k}^{\mmb}  \propto
\Xi_{\k}^{\mmb+\hbar {\kappa} }, 
\qquad \hat{B}_\k^- \Xi_{\k}^{\mmb+
\hbar{\kappa}}  \propto \Xi_{\k}^{\mmb}\, .
\end{equation}
In this way, by acting on the subspace of eigenfunctions $\Xi_{\k}^{\mmb}$ we find that
\begin{equation}\label{cqhb1} 
 \bigl[\hat\mm,\hat{B}_\k^\pm \bigr]=
\pm \hbar  {\kappa} \,\hat{B}_\k^\pm 
\quad
\Longleftrightarrow
\quad
\hat{B}_\k^\pm\, \hat\mm = (\hat\mm \mp \hbar\k)\hat{B}_\k^\pm \,.
\end{equation}
Therefore,
\begin{equation}\label{cqhb2}\nonumber
\bigl[  (\hat\mm)^2  ,\hat{B}_\k^\pm \bigr]= 
\hbar{\kappa} \left(\pm 2\hat\mm- 
\hbar{\kappa} \right)\hat{B}_\k^\pm \, ,
\end{equation}
that is, 
\begin{equation}\label{cqhb3}
\bigl[ \hat{H}_{\kappa}^{\xi},\hat{B}_\k^+ \bigr]= 
\hbar  \left(  \hat\mm- 
\frac 12 \hbar{\kappa}  \right)\hat{B}_\k^+ \, , \qquad
\bigl[ \hat{H}_{\kappa}^{\xi},\hat{B}_\k^- \bigr]= 
- \hat{B}_\k^-\,  \hbar \left(  \hat\mm- 
\frac 12 \hbar{\kappa}  \right)
\, .
\end{equation}
Likewise, we find that
\begin{equation}\label{qbpm}
\bigl[\hat{B}_{\k}^-,\hat{B}_{\k}^+\bigr]= 
\hbar\, \hat\mm \, .
\end{equation}
Note that the Lie brackets (\ref{cqhb3}) and (\ref{qbpm})  are just the quantum analogues of the Poisson algebra (\ref{commpt1}) and that a pure quantum-curvature term $\hbar\k/2$ arises, which is obviously negligible  at   both the classical curved and quantum flat (Euclidean) frameworks.

Finally, the commutation rules between the  operators $\hat{B}_\k^\pm$ and the complete Hamiltonian 
$\hat{H}_{\kappa}$  (\ref{hq1})  are shown to be
\begin{equation}\nonumber
[\hat{H}_{\kappa},\hat{B}_\k^\pm ]= \frac{\hbar\gamma^2}{\Ck_{\kappa}^2(y)}
\left(\pm \hat\mm- 
\frac 12 \hbar{\kappa}  \right)\hat{B}_\k^\pm  \, ,
\end{equation}
which can be compared with the Poisson algebra (\ref{commpt2}).

%%%%%%%%%%%%%%%%%%%%%%%%%%%%%%%%%%%%%%%%%%%%%%%%%%%%%%%%%5
\subsect{Shift operators for the Hamiltonian $\hat{H}_{\kappa}$}

By making use of the operator $\hat\mm$ (\ref{e}), the Hamiltonian  $\hat{H}_{\kappa}$  (\ref{hq1}), acting on the eigenfunctions (\ref{eve4}) can be rewritten  as 
\begin{equation}\label{hqy1}
\hat{H}_{\k}(\gamma \hat\mm)  = -\frac{\hbar^2}{2}\,\frac{ \partial^2}{\partial y^2}+\frac{\hbar^2}{2} \,\kappa \Tk_{\kappa}(y)\,\frac{ \partial}{\partial y}
+\frac{(\gamma \hat\mm)^2}{2\k\Ck_{\kappa}^2(y)}-
\frac{\omega^2}{2 \kappa} \,,
\end{equation}
where we have stressed its dependence on $\gamma\hat\mm$.
Now, it can be proven that $ \hat{H}_{\k}  $  can be factorized in   terms of two first-order differential operators plus a diagonal one, namely
\begin{equation}\label{fachqy3}
\hat{H}_{\k}  
=\hat{A}^+_{\k}\hat{A}^-_{\k}
+\hat\lambda^A_{\k},
\end{equation}
where
\begin{equation}
\begin{array}{l}
\displaystyle \hat{A}_{\k}^+ 
=-\frac{\hbar}{\sqrt{2}}\,  \frac{\partial}{\partial y}-   \frac 1{\sqrt{2}}  (\gamma\hat\mm-  \hbar{\kappa}  )  \Tk_{\kappa}(y),
\\[2.ex]
\displaystyle \hat{A}_{\k}^-
=\frac{\hbar}{\sqrt{2}}\,  \frac{\partial}{\partial y}-   \frac {\gamma\hat\mm}{\sqrt{2}}    \Tk_{\kappa}(y) ,\\[2.ex]
\displaystyle \hat\lambda^A_{\k}=\frac{\gamma\hat\mm}{2\kappa}(\gamma\hat\mm- \hbar{\kappa}   ) - \frac{\omega^2}{2\kappa} .
\nonumber
\end{array}
\ee
These expressions are worth to be compared with (\ref{cht2}) and (\ref{capm}).  
Then, the action of the shift operators on the eigenfunctions $Y_{\k}^{\gamma\mmb}$ can be shown to be of the type
\begin{equation}
\begin{array}{l}
  \hat{A}_\k^+ Y_{\k}^{\gamma\mmb-\hbar\k } 
  \propto Y_{\k}^{\gamma\mmb  } \,  ,
\qquad
  \hat{A}_\k^- Y_{\k}^{\gamma\mmb} \propto Y_{\k}^{\gamma\mmb -\hbar\k } \, .
\nonumber
\end{array}
\ee
Consequently,  these operators change the parameter $\gamma\mmb\to  \gamma\mmb \pm\hbar\k$ of the eigenfunctions, but keep the energy $E_{\kappa}$ constant. In this sense, they   are called {\em pure-shift }operators, and the following commutation relations can be derived
\be
 [\hat{H}_{\kappa}  ,\hat{A}_\k^+]
 =- {\hbar} \, \hat{A}_\k^+\left(\frac{2\gamma\hat\mm-\hbar  {\kappa} }{2\Ck_{\kappa}^2(y)}\right), \qquad [\hat{H}_{\kappa}  ,\hat{A}_\k^-]= {\hbar} \left(\frac{2\gamma\hat\mm- \hbar{\kappa} }{2\Ck_{\kappa}^2(y)}\right)\hat{A}_\k^- ,
 \nonumber
 \ee
where a quantum-curvature contribution $\hbar\k$ appears again. These commutation
relations are straightforwardly proven to be equivalent to the following so-called ``intertwining relations"
\begin{equation}
\hat{A}_\k^- \hat{H}_{\kappa}(\gamma\hat\mm)
= \hat{H}_{\kappa}(\gamma\hat\mm-\hbar\k) \hat{A}_\k^-\,,\qquad
\hat{A}_\k^+ \hat{H}_{\kappa}(\gamma\hat\mm-\hbar\k)
= \hat{H}_{\kappa}(\gamma\hat\mm) \hat{A}_\k^+\,.\label{intamp}
\end{equation}

%%%%%%%%%%%%%%%%%%%%%%%%%%%%%%%%%%%%%%%%%%%%%%%%%%%%%%%%%55
\subsect{Quantum symmetries}

So far we have all the ingredients to construct the ``additional"  symmetry operators $ \hat X_\k^{\pm}$ for the quantum Hamiltonian $\hat{H}_{\kappa}$ (\ref{hq3}) in the rational $\gamma=m/n$ case, that can be defined as
\begin{equation}
 \hat X_\k^{\pm}= (\hat A_\k^{\pm})^{m} (\hat B_\k^{\pm})^n,\qquad m,n\in \mathbb N^\ast .
 \label{ppc}
 \end{equation}
 
The proof that $[\hat H_\kappa, \hat X_\kappa^\pm]=0$ when $\gamma=m/n$ can be obtained by direct computation through the action on the subspace of eigenfunctions for $H_\kappa$ (see Appendix B for details).  The set of symmetries $(\hat H_\k, \hat H_\k^\xi,\hat X_\k^\pm)$ is not algebraically independent;
due to the factorization properties (\ref{factb}) and (\ref{fachqy3}),
the products $\hat X_\k^+\hat X_\k^-$ and $\hat X_\k^-\hat X_\k^+$ are
functions of $\hat H_\k, \hat H_\k^\xi$.
Similarly to the  classical case  (\ref{even}) and (\ref{odd}), 
we can define real polynomial quantum symmetries $\hat \xx_\k$ and $\hat \yy_\k$ of orders  $(m+n)$th  and $(m+n-1)$ in the momentum operators.
These sets of symmetries close a polynomial algebra 
(see \cite{lissajous} for more details).
 
All the above results are summarized as follows. 

\medskip

\noindent
{\bf Theorem 4.} {\em {\rm (i)} The quantum Hamiltonian $\hat H_\k$ (\ref{hc11}) defines  an integrable quantum  system for any value of the   parameter $\gamma$,  since it commutes with  the operator    $\hat H^\xi_\k$   (\ref{hq}).

\noindent
 {\rm (ii)} When $\gamma$ is a rational parameter,    $\hat H_\k$ defines   a  superintegrable  anisotropic quantum  curved oscillator  with    additional symmetry operators   given by (\ref{ppc}). The   sets $(\hat H_\k,\hat H_\k^\xi,  \hat X_\k^+)$ and $(\hat H_\k, \hat H_\k^\xi,  \hat X_\k^-)$ are formed by three algebraically  independent operators.
}

\medskip

 Let us illustrate this result through some particular systems.

%%%%%%%%%%%%%%%%%%%%%%
\subsection{Symmetries of the 1\,:\,1 case}

The simplest case of the quantum anisotropic curved oscillator corresponds to set
 $\gamma=1$ which, in fact, gives rise to the isotropic case. In this case we have the symmetry operators
\begin{equation}
\hat X_\k^{\pm}= (\hat A_\k^{\pm}) (\hat B_\k^{\pm}) 
= \pm \hat \yy_\k \hat\mm + \hat \xx_\k ,
\nonumber
\end{equation}
where $\hat \xx_\k$ is a polynomial symmetry of degree two, 
while $\hat \yy_\k$ has degree one
in the momentum operators.
These symmetries take the explicit form
\begin{eqnarray}
&&\hat \xx_\k
=\frac{\hbar^2}2  \Ck_\k(\xi) \partial_{\xi }\partial_y  - \frac 12 \Sk_\k(\xi) \Tk_\k(y)  (\hat\mm)^2,
\nonumber\\
&&
\hat\yy_\k = -\frac \hbar 2\bigl(  \Sk_\k(\xi)   \partial_y - \Ck_\k(\xi) \Tk_\k(y)  \partial_\xi \bigr) ,
\nonumber
\end{eqnarray}
where $\xi=x$. Recall that, according to (\ref{eve3}) and (\ref{e}), we can replace  $(\hat\mm)^2$ by 
$2\k \hat{H}_{\k}^{\xi}$.

If we take the classical limit $\hbar\to 0$ (so $-i\hbar\partial_\xi\to p_\xi$ and $\hat\mm\to \mm$),  we recover
  the classical symmetries on the curved spaces given in (\ref{sa}):  $\hat X_\k\to X_\k$ and $\hat Y_\k  \to i Y_\k$.
Furthermore, in the limit $\k\to 0$, they become the classical
Euclidean expressions shown in (\ref{al}).

%%%%%%%%%%%%%%%%%%%%%%
\subsection{Symmetries of the 2\,:\,1 case}

When $\gamma=2$ the symmetries come from the operators
\begin{equation}
\hat X_\k^{\pm}= (\hat A_\k^{\pm})^2 (\hat B_\k^{\pm}) 
= \hat \xx_\k \hat\mm \pm \hat \yy_\k ,
\nonumber
\end{equation}
where $\hat \xx_\k$ is a polynomial symmetry of degree two, while $\hat \yy_\k$ has degree three in the momentum operators.
Explicitly, we have
\begin{eqnarray}
&&\hat \xx_\k
= \frac 2{\sqrt{2}} \Sk_\k(\xi) \Tk^2_\k(y)  (\hat\mm)^2
+ \frac{\hbar^2 \k}{2\sqrt{2}}   \Sk_\k(\xi) \Tk_\k(y) \partial_y +\frac{ \hbar^2}{2\sqrt{2}} \Sk_\k(\xi)\partial^2_{y}  \nonumber
\\[1.5ex]
&&\qquad\quad 
-\frac{\hbar^2}{\sqrt{2}}  \Ck_\k(\xi) \left( \frac1{\Ck^2_\k(y)}    +\k \Tk^2_\k(y)  \right)  \partial_\xi
-\frac{2  \hbar^2}{\sqrt{2}}  \Ck_\k(\xi)\Tk_\k(y)\partial_{\xi}\partial_y \, ,
\nonumber\\[2.ex]
&&
\hat \yy_\k = \frac{2\hbar}{\sqrt{2}}
\left(  \Sk_\k(\xi) \left(    \frac1{\Ck^2_\k(y)} -\frac 12 \right) +  \Sk_\k(\xi)  \Tk_\k(y)
  \partial_y-  \Ck_\k(\xi) \Tk^2_\k(y) \partial_\xi
\right) (\hat\mm)^2          \nonumber
\\[1.5ex]
&&\qquad \quad 
-\frac{ \hbar^3\k}{2\sqrt{2}} \Ck_\k(\xi)\Tk_\k(y)\partial_{\xi}\partial_y - \frac{\hbar^3}{2\sqrt{2}}\Ck_\k(\xi)\partial_{\xi}\partial^2_y \, .
\nonumber
\end{eqnarray}

In the   limit $\hbar\to 0$, these symmetries turn
into the   classical counterparts of (\ref{pol21aclas}) and (\ref{pol21bclas}): 
$\hat X_\k \to X_\k$ and $\hat Y_\k\to i Y_\k$; recall that   now $\xi=2x$.
If on this latter result we take the flat limit $\k\to 0$ we recover  the Euclidean constants of motion of (\ref{am}).

%%%%%%%%%%%%%%%%%%%%%%
\subsection{Symmetries of the 1\,:\,2 case}

If we set $\gamma=1/2$, the symmetries read
$$
\hat X_\k^{\pm}= (\hat A_\k^{\pm}) (\hat B_\k^{\pm})^2 
= \hat \xx_\k \hat\mm \pm \hat \yy_\k \,,
$$
where, as in the previous case,  $\hat \xx_\k$ is a  second-order symmetry while    $\hat \yy_\k$ is  a third-order one.
These operators take the following form
\begin{eqnarray}
&&\hat \xx_\k
= - \frac1{4\sqrt2} \Sk^2_\k(\xi) \Tk_\k(y)  (\hat\mm)^2 
- \frac{\hbar^2}{4\sqrt2}    \Ck^2_\k(\xi) \Tk_\k(y)  \partial^2_{\xi}   \nonumber
\\[1.5ex]
&&\qquad\quad
+\frac{\hbar^2}{4\sqrt2} \bigl(2 \Ck_\k(2 \xi) \partial_y
+ \Sk_\k(2\xi)(\k \Tk_\k(y) \partial_\xi + 2 \partial_{\xi}\partial_{y}) \bigr),
\nonumber\\[2.ex]
&&
\hat \yy_\k = 
\frac{\hbar}{4\sqrt2} \left(\Tk_\k(y)(  \Ck_\k(2\xi)+\Sk_\k(2\xi) \partial_\xi) -2 \Sk^2_\k(\xi)\partial_y 
\right) (\hat\mm)^2          \nonumber
\\[1.5ex]
&&\qquad \quad
+ \frac{\hbar^3}{2\sqrt2}  \Ck_\k(\xi) \bigl( 2 \k \Sk_\k(\xi)\partial_{\xi}\partial_{y} 
-\Ck_\k(\xi)\partial_{\xi}^2 \partial_{y} \bigr) \, .
\nonumber
\end{eqnarray}
 
 Under the     limit $\hbar\to 0$, these symmetries give rise to  the curved classical functions   (\ref{clas12a}) and (\ref{clas12b}) provided that   $\xi = x/2$.

In the same manner, other    $m:n$ quantum  oscillators  can straightforwardly 
be worked out, and, obviously, the expressions for their symmetries    become rather cumbersome.

%%%%%%%%%%%%%%%%%%%%%%%%%%%%%%%%%%%%%%%%%%%%%%%%%%%%%%%%%55

\sect{Spectrum of the anisotropic oscillator on the sphere}

Due to the different properties of the spectra in the $\k>0$ and the $\k<0$ cases, both quantum systems are worth to be analysed separately. Let us firstly consider the anisotropic oscillator on the sphere ${\mathbf S}^2$ with arbitrary positive curvature $\k>0$. Take the eigenfunctions 
$\Psi_\k^E= \Xi_\k^\mmb\,Y_{\k}^{\gamma \mmb} $, given in terms of the eigenfunctions
of the one-dimensional Hamiltonians (\ref{eve3}) and (\ref{eve4}). The eigenvalue
equation (\ref{eve3})
$\hat{H}_{\kappa}^{\xi} \,\Xi_\k^\mmb(\xi)=E_{\kappa}^{\xi}\,\Xi_\k^\mmb(\xi)$,  
for $\Xi_\k^\mmb$, corresponds to the well-known
trigonometric P\"oschl--Teller Hamiltonian \cite{kuru09,kuru12} whose eigenvalues are given
by
\bea
&&E_\k^{\xi}\equiv E_{\k,\mmc}^{\mu}=\frac 1{2\k}  \bigl( \mmc+ (\mu+1)\hbar\k \bigr)^2  \label{eigenvaluexis}
\\[2pt]
&&\quad\  = \frac {\hbar}4 (1+2\mu) \sqrt{\hbar^2\k^2+ \frac{4 \omega^2}{\gamma^2 }}+ \frac{\hbar^2\k}{4}\left(1+2\mu+2\mu^2 \right)+\frac{\omega^2}{ 2\k\gamma^2} 
, \quad  \mu=0,1,2,\dots 
\nonumber
\eea
where the positive parameter $\mmc$ is given by
\be
\mmc(\mmc+\hbar\k)=\frac {\omega^2}{\gamma^2},\qquad \mmc= \frac 12 \left(\sqrt{\hbar^2\k^2+ 4 \omega^2/\gamma^2}-\hbar\k  \right) .
\label{chi}
\ee
Having in mind   (\ref{eve3}), we will also use the parameter $\mmb_\mu$ corresponding to such eigenvalues:
\be
\mmb_\mu=  \sqrt{2\k E_\k^{\xi}}=  \mmc+ (\mu+1)\hbar\k \,.
\label{energia2}
\ee

On the other hand, the eigenvalue equation for the Hamiltonian (\ref{Hk}),
$\hat{H}_{\kappa}Y_\k^{\gamma\mmb}(y)
=E_{\kappa} Y_\k^{\gamma\mmb}(y)$, satisfied by the second 
function $Y_\k^{\gamma \mmb}$, also corresponds  to a modified 
trigonometric P\"oschl--Teller  Hamiltonian with eigenvalues
\begin{equation}\nonumber
E_{\k,\gamma\mmb}^{\nu}
=\frac 1{2\k} \left( \gamma\mmb+\nu \hbar\k\right) \bigl(\gamma\mmb+(\nu+1) \hbar\k \bigr) -\frac{\omega^2}{2\k},\quad \nu=0,1,2,\dots
\end{equation}

Therefore, by replacing $\mmb$ in this expression  by the values $\mmb_\mu$ above obtained in (\ref{energia2}), 
we get the energy eigenvalues of the whole two-dimensional Hamiltonian (\ref{hq3}), namely
\bea
 &&\!\!\!\! \!\!\!\!\!\!\!\! \!\!\!\!
E_\k\equiv E_\k^{\mu,\nu}=\frac 1{2\k}\bigl(\gamma \mmc+\left[ \gamma(\mu+1)+\nu\right] \hbar\k  \bigr)  \bigl(\gamma \mmc+\left[ \gamma(\mu+1)+\nu+1\right] \hbar\k  \bigr)  -\frac{\omega^2}{2\k}  
\nonumber
\\[2pt]
&&\!\!\!\! \!\!\!\!
 =   {\gamma^2}  \left ( E_{\k,\mmc}^{\mu} -\frac{\omega^2}{2\k\gamma^2}\right)  +\frac{\hbar\gamma}2  \bigl( \mmc+ (\mu+1)\hbar\k \bigr)(2\nu+1)+\frac{\hbar^2\k}2\nu(\nu+1) .
\label{eigenvalues}
\eea
It is worth stressing that in this expression the role of the curvature is essential, since  it gives rise
to a quadratic dependence in terms of the quantum numbers $\mu, \nu$ instead of the linear Euclidean one.
The corresponding eigenfunctions (\ref{eve1})  take the form 
\[
\Psi^{\mu,\nu}_\k(\xi,y)
=\Xi^{\epsilon_\mu}_{\k}(\xi) Y_{\k,\nu}^{\gamma\mmb_\mu}(y)\,.
\]
According to (\ref{eigenvalues}), two of these eigenfunctions $\Psi_\k^{\mu,\nu}(\xi,y)$  and $\Psi_\k^{\mu',\nu'}(\xi,y)$ will have the 
same energy $E_\k^{\mu,\nu}=E_\k^{\mu',\nu'}$  if 
\be
\gamma(\mu'-\mu)+\nu'-\nu =0,
\label{const}
\ee
which is only satisfied when $\gamma=m/n$ with $m,n\in \mathbb N^\ast$, and these states are connected by the symmetry operators (\ref{ppc}). Therefore, we can conclude that:
\medskip

\noindent
{\bf Theorem 5.} {\em {\rm (i)} The spectrum of the quantum Hamiltonian $\hat H_\k$ (\ref{hq3}) on the sphere with $\k>0$  is given, for any value of the parameter $\gamma$, by 
(\ref{eigenvalues}) where the parameter $\mmc$ is written in (\ref{chi}).

\noindent
 {\rm (ii)} When $\gamma=m/n$ with $m,n\in \mathbb N^\ast$,  the spectrum (\ref{eigenvalues}) is degenerate, and the degeneracy is the same as in the Euclidean case.
 }

\medskip

Some comments on the Euclidean limit of the above results seem to be pertinent.
Although most of the computations have been carried out for $\k\ne 0$, the flat limit $\k\to 0$ can adequately be    performed on the final results.
Explicitly, the limit $\k\to 0$ of  the parameters $\mmc$   (\ref{chi})  and    $\mmb_\mu$   (\ref{energia2})  as well as  of the spectrum $E_\k^{\xi}\equiv E_{\k,\mmc}^{\mu}$ (\ref{eigenvaluexis}) is achieved as
\be
\lim_{\k\to 0}\mmc=\lim_{\k\to 0}\mmb_\mu=\frac{\omega}{\gamma} ,\qquad \lim_{\k\to 0}\left( E_{\k,\mmc}^{\mu}- \frac{\omega^2}{ 2\k\gamma^2}\right)= \frac{\hbar\omega}{2 \gamma}+  \mu\,   \frac{\hbar\omega}{ \gamma} ,
\label{limita}
\ee
such that the latter is just the spectrum ${E}^{\xi,\mu}$ (\ref{eigenvaluexyp}) of the one-dimensional  quantum Euclidean  Hamiltonian $\hat{H}^{\xi} $ (\ref{hqyp}).
And the complete spectrum $E_\k\equiv E_\k^{\mu,\nu}$ (\ref{eigenvalues})  directly reduces  to ${E}^{\mu,\nu}$ (\ref{eigenvaluep})  corresponding to 
the two-dimensional quantum Euclidean  Hamiltonian $\hat H$    (\ref{hq3p2}):
\be
\lim_{\k\to 0} E_\k^{\mu,\nu}= \hbar \omega\left( \tfrac 12 (\gamma+1) +\gamma \mu+\nu    \right)  \equiv {E}^{\mu,\nu} .
\label{limitb}
\ee

 %%%%%%%%%%%%%%%%%%%%%%%%%%%%%%%%%%%%%%%%%%%%%%%%%%%%%%%%%55

\sect{Spectrum of the anisotropic oscillator on the hyperboloid}

Now, we consider the hyperbolic space $ {\mathbf H}^2$ with arbitrary  negative curvature $\k=-\mk$  and we follow the same approach as in the sphere  $ {\mathbf S}^2$ with $\k>0$. We anticipate that,  although the same algebraic method holds in both spaces, the solution to the  eigenvalue problem   is quite different.

Let us consider the Hamiltonian 
$\hat{H}_\k \equiv  \hat{H}_{-\mk}$    
(\ref{hqy1}), where the factorized solutions take the form
$\Psi_{-|\k|}^E= \Xi_{-\mk}^\mmb\,Y_{-\mk}^{\gamma \mmb} $ in the same
way as in the previous section. Now, the eigenvalue
equation (\ref{eve3})
$\hat{H}_{-\mk}^{\xi} \,\Xi_{-\mk}^\mmb(\xi)=E_{-\mk}^{\xi}\,\Xi_{-\mk}^\mmb(\xi)$,  for $\Xi_{-\mk}^\mmb$, corresponds to the 
hyperbolic P\"oschl--Teller Hamiltonian \cite{kuru09,kuru12} whose eigenvalues are 
\bea
&& E_{-\mk}^{\xi}\equiv E_{\k,\mmc}^{\mu}= -\frac 1{2\mk}  \left( \mmc- (\mu+1)\hbar\mk \right)^2  \nonumber 
\\[2pt]
&&\ \  = \frac {\hbar}4 (1+2\mu) \sqrt{\hbar^2\mk^2+ \frac{4 \omega^2}{\gamma^2 }}- \frac{\hbar^2\mk}{4}\left(1+2\mu+2\mu^2 \right)-\frac{\omega^2}{ 2\mk\gamma^2} 
\nonumber
, \quad  \mu=0,1,2,\dots, \mu_{\max} ,
\eea
where the parameter $\mmc>\hbar\mk$ is given by
\be
\mmc(\mmc-\hbar\mk)=\frac {\omega^2}{\gamma^2},\qquad \mmc= \frac 12 \left(\sqrt{\hbar^2\mk^2+ 4 \omega^2/\gamma^2}+\hbar\mk  \right) .
\label{chib}
\ee
Therefore, there is only a finite number of bounded states. The value
$\mu_{\max}$ is the maximum integer such that   
 \be
\mu_{\max}< \frac{\mmc}{\hbar\mk} -1 \,.
\label{mua}
\ee
Again, instead of $E_{-\mk}^{\xi}$ we will   use the   parameter (\ref{eve3}),
\be
\mmb_\mu=  \sqrt{-2\mk E_{-\mk}^{\xi}}=  \mmc- (\mu+1)\hbar\mk \,.
\label{energia3}
\ee

The eigenvalue equation (\ref{eve4}),
$\hat{H}_{-\mk}Y_{-\mk}^{\gamma\mmb}(y)
=E_{-\mk} Y_{-\mk}^{\gamma\mmb}(y)$, satisfied by the  
function $Y_{-\mk}^{\gamma \mmb}$, is the one of a modified 
hyperbolic P\"oschl--Teller  Hamiltonian with eigenvalues
\begin{equation}\label{eigenvalueysb}
E_{-\mk,\gamma\mmb}^{\nu}=-\frac 1{2\mk} \left( \gamma\mmb-\nu \hbar\mk\right) \bigl(\gamma\mmb-(\nu+1) \hbar\mk \bigr) +\frac{\omega^2}{2\mk},\quad \nu=0,1,2,\dots, \nu_{\max}.
\end{equation}
Such a maximum value $\nu_{\max}$ of  the quantum number $\nu$ is the maximum integer
such that  
\be
\nu_{\max}<
\frac{\gamma\mmb}{\hbar \mk}-1 \, .
\label{nua}
\ee
Next,  replacing  $\mmb$ in (\ref{eigenvalueysb}) by the values $\mmb_\mu$   obtained in (\ref{energia3}), 
we obtain the energy eigenvalues of the quantum Hamiltonian $\hat{H}_\k \equiv  \hat{H}_{-\mk}$ (\ref{hq3}):
\bea
&&\!\!\!\!\!\!\!
E_{-\mk}\equiv E_{-\mk}^{\mu,\nu}=-\frac 1{2\mk}\bigl(\gamma \mmc-\left[ \gamma(\mu+1)+\nu\right] \hbar\mk  \bigr)  \bigl(\gamma \mmc-\left[ \gamma(\mu+1)+\nu+1\right] \hbar\mk  \bigr) +\frac{\omega^2}{2\mk} 
\nonumber\\[2pt]
&& \quad\  =   {\gamma^2}  \left ( E_{\k,\mmc}^{\mu}+\frac{\omega^2}{2\mk\gamma^2}\right)  +\frac{\hbar\gamma}2  \bigl( \mmc- (\mu+1)\hbar\mk \bigr)(2\nu+1)-\frac{\hbar^2\mk}2\nu(\nu+1) .
\label{eigenvalues3}
\eea
Clearly,  the number of  eigenvalues is finite because of the constraints on $\mu$  and  $\nu$, that is,
  we have a fixed value of $\mmc$ (\ref{chib}), which determines $\mu_{\max}$ (\ref{mua}) and, then, for any
allowed value of $\mu$ there is a maximum value for $\nu$, $\nu_{\max}(\mu)$ determined by (\ref{nua}). Consequently, on $ {\mathbf H}^2$ there is a finite number of bound states in contradistinction  with the $ {\mathbf S}^2$ case.
The corresponding eigenfunctions are written in the form 
$$
\Psi_{-\mk}^{\mu,\nu}(\xi,y)=
\Xi_{-\mk}^{\mmb_\mu}(\xi)\, Y_{-\mk, \nu}^{\gamma\mmb_\mu}(y) \,.
$$

The degeneracy of the energy levels can  be discussed in a similar
way as on $\mathbf{S}^2$: $\Psi_{-\mk}^{\mu,\nu}(\xi,y)$  and $\Psi_{-\mk}^{\mu',\nu'}(\xi,y)$ will 
have the same energy $E_{-\mk}^{\mu,\nu}$ whenever the constraint (\ref{const}) is fulfilled. This  means that the degeneration can take place
only when $\gamma=m/n$ for  $m,n\in \mathbb N^\ast$.

 Summing up, we have shown that:
\medskip

\noindent
{\bf Theorem 6.} {\em {\rm (i)} The spectrum of the quantum Hamiltonian $\hat H_{-\mk}$ (\ref{hq3}) on the hyperbolic space with $\k=-\mk<0$  is given, for any value of the parameter $\gamma$, by a finite number of eigenvalues
(\ref{eigenvalues3}) where the parameter $\mmc$ is written in (\ref{chib}).

\noindent
 {\rm (ii)} When $\gamma=m/n$ with $m,n\in \mathbb N^\ast$,  the spectrum (\ref{eigenvalues3}) is degenerate, and the degeneracy is the same as in the Euclidean case.
}

\medskip   

  The Euclidean limit of the above results can be easily obtained in the same way as in (\ref{limita}) and (\ref{limitb}). Finally, we should mention that
for negative curvature $\k = - \mk$, there exist unbounded states which may
come from any of the one-dimensional hyperbolic P\"oschl--Teller potentials
which take part in the total Hamiltonian.
This feature, of course, is not present for the anisotropic oscillator on $ {\mathbf S}^2$.

%%%%%%%%%%%%%%%%%%%%%%%%%%%%%%%%%%%%%%%%%%%%%%%%%

\sect{Concluding remarks }

In this work we have identified the form of the generic classical and quantum anisotropic oscillator 
system on the sphere and the hyperboloid, in such a way that this new Hamiltonian 
keeps the integrability properties of the known Euclidean anisotropic oscillator for any value of the frequencies.
Moreover, when the frequencies are commensurate, superintegrability arises and, in particular, the previously known curved anisotropic oscillators which correspond to the ratio $1\!:\!1$ or $2\!:\!1$ are recovered.

There are two key points in the approach here presented. 
The first one consists in a formulation  depending on the curvature parameter $\k$, in such a way that the algebraic treatment 
is simultaneous for both the sphere and the hyperboloid. At the same time,
this viewpoint allows us to get the Euclidean expressions  in the limit $\k\to 0$, thus showing explicitly
that our systems are indeed curved integrable deformations of the Euclidean anisotropic oscillator. 
Of course, there are also many
properties that depend on the sign of $\k$, and they have to be treated separately. For instance, the spectrum of the quantum
anisotropic oscillator on ${\bf S}^2$ is purely discrete (and has infinite values), whilst  
a (finite) discrete spectrum plus a continuous one arises for the system on ${\bf H}^2$,
as it has been explicitly discussed.

The second key point is the choice of geodesic parallel coordinates on the curved 
surfaces in order to get the simplest possible expression of the corresponding anisotropic oscillators. These
coordinates turn out to be  the appropriate curved analogue of the Euclidean Cartesian
coordinates, which are used to write the planar anisotropic oscillator in a separable manner.

In order to get a unified approach to find
the symmetries for both, the classical and quantum systems, we have  applied a factorization approach
\cite{Kuru1,Kuru3,lissajous,Ragnisco1}. This method turns out to be helpful in order to highlight the correspondence between the classical and quantum algebraic symmetries. In this respect, it is worth to stress that in the quantum context
we have kept the quantum constant $\hbar$ in all the expressions. This is quite relevant
when we compare  quantum and classical results through the limit
$\hbar\to 0$ (together with other considerations). 

We also recall that other methods have also been
designed to deal with the symmetries of this kind of systems; for instance we can mention an action-angle
(or Hamilton--Jacobi) based procedure
 considered for the classical systems in \cite{miller12} and
the type of recurrence
relation arguments used in the quantum counterparts \cite{miller11}. Similar approaches
for a family of superintegrable models were applied in \cite{post}, a coalgebra procedure has been developed in \cite{riglioni13}, while  other different viewpoints/approaches can be found in~\cite{Marquette1, Marquette2, gonera12,hakobyan12,rastelli15}.

There are several open problems related to anisotropic curved oscillators,
which are worth to be investigated in the near future. For instance: 
(i) The construction of the anisotropic curved oscillators in three (and $N$) dimensions.
(ii) The formulation of anisotropic oscillators on $(1+1)$  relativistic spacetimes: AdS, dS and Minkowski~\cite{ballesteros} (see also~\cite{so221, so22} for the   oscillator  problem on the $SO(2,2)$ hyperboloid).
(iii) The generalization of the so-called caged anisotropic oscillator studied in~\cite{Verrier} to curved spaces by adding centrifugal terms \cite{ballesteros13,Letter, Montreal}. 

Finally, we remark that in order to keep the length of this paper under reasonable limits, we have not included the full study of the algebra generated by the symmetries in the generic case. However, this structure
can straightforwardly be derived by following the procedure shown in~\cite{lissajous}.

%%%%%%%%%%%%%%%%%%%%%%%%%%%%%%%%%%%%%%%%%%55

%%%%%%%%%%%%%%%%%%%%%%%%%
\section*{Appendix A}
\setcounter{equation}{0}
\renewcommand{\theequation}{A.\arabic{equation}}

The three Riemannian  spaces $ {\mathbf S}^2$, $ {\mathbf E}^2$ and $ {\mathbf H}^2$ can be simultaneously studied by considering that their constant Gaussian curvature $\k=\pm 1/R^2$  plays the role of a deformation/contraction parameter (see~\cite{mariano99,anisotropic,conf, Letter, Montreal} and references therein). These three spaces can be       embedded  in the  linear space $\mathbb R^{3}$ with
  ambient  or Weierstrass coordinates $(x_0,x_1,x_2)$ subjected to  the constraint 
   \begin{equation}
\Sigma_\k\equiv x_0^2+\k \left(x_1^2+   x_2^2\right)  =1 ,
\label{ca}
\end{equation}
 such that the origin corresponds to   the point $O=(1,0,0)\in \mathbb R^{3}$. 
If   $\k=1/R^2>0$, we recover the sphere $ {\mathbf S}^2$ and when $\k=-1/R^2<0$, we find the two-sheeted hyperboloid. The null curvature case can be understood as a flat contraction $\k= 0$  (i.e.~the limit $R\to \infty$),  giving rise to   two Euclidean planes $x_0=\pm 1$ with Cartesian coordinates $(x_1,x_2)$. We shall identify the hyperbolic space  $ {\mathbf H}^2$ with  the upper sheet of the hyperboloid with $x_0\ge 1$ and the Euclidean space $ {\mathbf E}^2$   with the plane $x_0=+1$. Although   $\k\ne 0$ can  always be scaled to $\pm 1$,   the explicit presence of the curvature parameter will  make evident all the deformation processes from the Euclidean systems to the curved ones and, conversely,  the contraction from the latter to the former ones. 
 
In this aproach, the metric on the curved spaces comes from  the usual metric in $\mathbb R^{3}$ divided by the curvature $\k$ and
restricted to $\Sigma_\k$ (\ref{ca}):
 \begin{equation}
{\rm d} s^2=\left.\frac {1}{\k}
\left({\rm d} x_0^2+   \k \left( {\rm d} x_1^2+   {\rm d} x_2^2 \right)
\right)\right|_{\Sigma_\k}  =    \frac{\k\left(x_1{\rm d} x_1 + x_2{\rm d} x_2 \right)^2}{1-  \k \left(x_1^2+   x_2^2\right)}+  {\rm d} x_1^2+   {\rm d} x_2^2 .
\label{ce}
\end{equation}

The ambient coordinates $(x_0,x_1,x_2)$ can be parametrized in terms of two intrinsic variables in different ways (see, e.g.,~\cite{anisotropic}). In particular, we can consider geodesic polar $(r,\te)$ and geodesic parallel coordinates $(x,y)$ (see Fig.~1). The parametrization of the ambient coordinates  in the   variables $(x,y)$   and  $(r,\te)$, so  fulfilling the constraint (\ref{ca}), reads~\cite{mariano99,conf, Letter, Montreal}  
\bea
&& x_0=\Ck_\kk(x)\Ck_\kk(y)=\Ck_\kk(r),\nonumber\\[2pt]
&& x_1=\Sk_\kk(x)\Ck_\kk(y)=\Sk_\kk(r)\cos\te , \label{cf}\\[2pt]
&& x_2=\Sk_\kk(y)=\Sk_\kk(r)\sin\te,
\nonumber
\eea
which on $ {\mathbf E}^2$ with $\k=0$ reduce to
$$
 x_0=1,\qquad  x_1=x= r\cos\te,    \qquad x_2= y =r\sin\te.
 $$

In the three spaces,  the coordinates $x$, $y$ and $r$  have dimensions of length, while $\phi\in[0,2\pi)$ is always an ordinary angle. However,   on $ {\mathbf S}^2$ with $\k=1/R^2$,  the dimensionless variable $r/R$    is usually considered  instead of $r$. All these coordinates are represented  in Fig.~3.

%%%%%%%%%%%%%%%% figure 3%%%%%%%%%%%%%%%%%%%%%%%%%%%%%

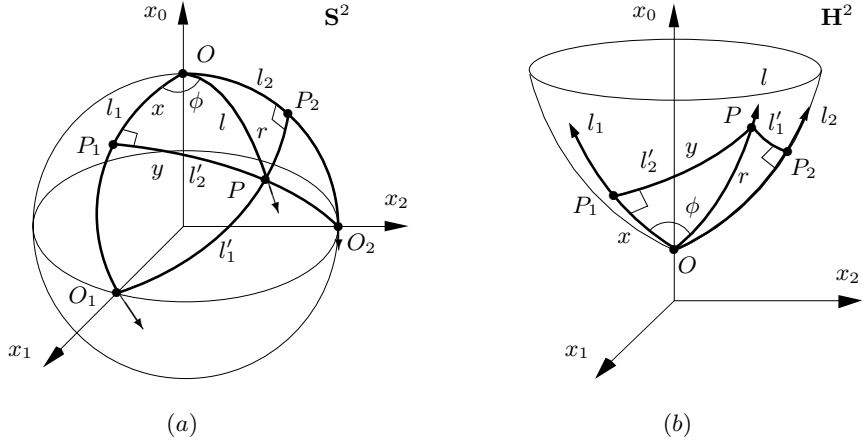
\begin{figure}[t]
\setlength{\unitlength}{1mm}
\begin{picture}(120,60)(0,0)
\footnotesize{
\put(14,8){\includegraphics[scale=1]{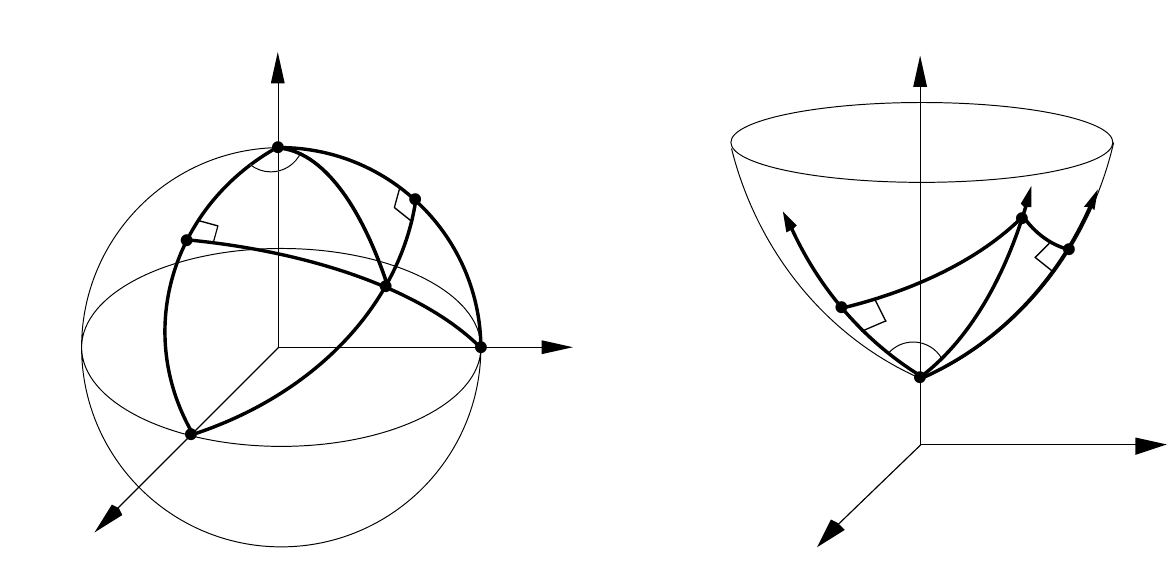}}
\put(37,58){$x_0$}
\put(19,13){{$x_1$}}
\put(69,33){$x_2$}
 \put(44,52){$O$}
\put(27,20){{$O_1$}}
\put(64,27){$O_2$}
 \put(43,36){$l'_2$}
 \put(38,37){$y$}
 \put(48,34){$P$}
 \put(38,45){$x$}
\put(47,26){$l'_1$}
 \put(57,46){$P_2$}
 \put(28,40){$P_1$}
  \put(52,49){$l_2$}
 \put(32,45){$l_1$}
\put(43,46){$\phi$}
 \put(47,43){$l$}
 \put(52,42){$r$}
  \put(102,58){$x_0$}
\put(93,13){{$x_1$}}
\put(129,23){$x_2$}
\put(119,49){$l$}
\put(114,44){$P$}
\put(96,43){$l_1$}
\put(127,44){$l_2$}
\put(103,38){$l'_2$}
\put(108,24){$O$}
\put(123,37){$P_2$}
\put(94,32){$P_1$}
\put(116,36){$r$}
\put(109,32){$\phi$}
\put(100,28){$x$}
\put(120,43){$l'_1$}
\put(109,40){$y$}
\put(61,57){${\bf S}^2$}
\put(127,57){${\bf H}^2$}
\put(40,3){$(a)$}
\put(106,3){$(b)$
\put(-47,27){\vector(0,-1){3}}
\put(-57,35){\vector(1,-3){2}}
\put(-76,18){\vector(2,-3){3}}
}
}
\end{picture}
\caption{\footnotesize Ambient   $(x_0,x_1,x_2)$,  geodesic parallel $(x,y)$ and geodesic polar $(r,\phi)$ coordinates. The origin is $O=(1,0,0)$. $(a)$ On the sphere  $ {\mathbf S}^2$ with $\k=+1$; $O_1=(0,1,0)$ and $O_2=(0,0,1)$.
  $(b)$ On the hyperboloid  with   $\k=-1$ and $x_0\ge 1$.  }
\end{figure}

%%%%%%%%%%%%%%%%%%%%%%%%%%%%%%%%%%%%%%%%%%%%%%%%%%%%%%

By introducing (\ref{cf}) in the metric (\ref{ce}) we find the metrics
\be
{\rm d} s^2=\Ck^2_\k(y){\rm d} x^2 + {\rm d} y^2   =      {\rm d} r^2+  \Sk^2_\k(r)  {\rm d} \phi^2 .
\label{fa}
\ee
The kinetic energy Lagrangian for a free particle moving on these spaces can be straightforwardly derived from~\eqref{fa}. Explicitly, let $(p_x,p_y)$  and $(p_r,p_\te)$ be, in this order, the conjugate momenta for $(x,y)$ and $(r,\te)$. Then, the free kinetic energy Hamiltonian ${\cal T}_\k$ that generates the geodesic dynamics in the curved space is given by
\be
 {\cal T}_\k =\frac 12 \left(\frac{p_x^2}{\Ck^2_\k(y)}  +  p_y^2\right) =\frac 12 \left( p_r^2+\frac{p_\te^2}{\Sk^2_\kk(r)} \right) .
 \label{ci}
 \ee
Indeed, when $\k= 0$  we recover the   Euclidean kinetic energy.
$$
 {\cal T}_0 =\frac 12 \left( {p_x^2} +  p_y^2\right) =\frac 12 \left( p_r^2+\frac{p_\te^2}{r^2} \right) .
 \label{cj}
$$
In order to guarantee that (\ref{ci}) is well defined,  we find that the coordinates $(x,y)$   and  $r$ have to be defined in the following  intervals:
\bea
 {\mathbf S}^2\ (\k>0) :&&      -\frac{\pi}{\sqrt{\k}}< x\le \frac{\pi}{\sqrt{\k}} ,\quad   -\frac{\pi}{2\sqrt{\k}}< y< \frac{\pi}{2\sqrt{\k}} ,  \quad  0< r< \frac{\pi}{\sqrt{\k}}  .   \label{zzbb}\\[2pt]
 {\mathbf H}^2\ (\k<0) :&&      -\infty< x< \infty ,\quad   -\infty< y< \infty,  \quad  0< r<\infty  .
 \label{zb}
\eea

%%%%%%%%%%%%%%%%%%%%%%%%%%%%%%%%%%%%%%%%%%%%%%%%%%%%%%%%%%%%%%%%

\section*{Appendix B}
\setcounter{equation}{0}
\renewcommand{\theequation}{B.\arabic{equation}}

In order to check that  $ \hat X_\k^{\pm}$ are a pair of symmetries of the Hamiltonian $\hat{H}_{\kappa}$ (\ref{hq3}) when the coefficient $\gamma$ takes the rational value, let us use the following notation
for the Hamiltonian  (\ref{hq1})
\[
\hat{H}_{\kappa}
= -\frac{\hbar^2}{2}\,\frac{ \partial^2}{\partial y^2}
+\frac{\hbar^2}{2} \kappa \Tk_{\kappa}(y)\frac{ \partial}{\partial y}
+\frac{(\gamma \hat\mm)^2}{2\k \Ck_{\kappa}^2(y)}
-\frac{\omega^2}{2 \kappa} 
\equiv\hat{H}_{\kappa}\bigl(\gamma  \hat\mm \bigr),
\]
where we recall the definition (\ref{e}) of the operator $\hat \mm$.  
Let us consider, for instance, the product 
$\hat X_\k^{+} \hat{H}_{\kappa}(\gamma \hat\mm)$ and let us move the operators 
$\hat X_\k^{+}$ to the r.h.s.~of the Hamiltonian. First, by making
use of (\ref{cqhb1}) we move the $\hat B_\k^{+}$ operators:
\[
\hat X_\k^{+} \hat{H}_{\kappa}(\gamma \hat\mm)
=(\hat A_\k^{+})^{m} (\hat B_\k^{+})^n\hat{H}_{\kappa}(\gamma \hat\mm)
= (\hat A_\k^{+})^{m}\hat{H}_{\kappa} 
\bigl(\gamma (\hat\mm-n\hbar\k) \bigr)
(\hat B_\k^{+})^n \, .
\]
Next, we translate the $\hat A_\k^{+}$ operators to the  r.h.s.~by means of the
intertwining (\ref{intamp}) and require that we should obtain
$\hat{H}_{\kappa}(\gamma \hat\mm) \hat X_\k^{+} $, that is, 
\[
\hat X_\k^{+} \hat{H}_{\kappa}(\gamma \hat\mm)=  
\hat{H}_{\kappa}\bigl(\gamma (\hat\mm-n\hbar\k)+m\hbar\k \bigr) 
(\hat A_\k^{+})^{m}(\hat B_\k^{+})^n 
=\hat{H}_{\kappa}(\gamma \hat\mm) \hat X_\k^{+} 
\,.
\]
Therefore, we get the condition
\begin{equation}\nonumber
\gamma \hat\mm=\gamma (\hat\mm-n\hbar\k)+m\hbar\k\quad  \Longleftrightarrow\quad  \gamma= m/n
\end{equation}
and in that case $\hat X_\k^{+}$ (and $\hat X_\k^{-}$ as well)  will be a symmetry operator of 
the Hamiltonian. Notice that the operator products in (\ref{ppc}) 
are written in terms of the operator 
$\hat \mm$. In order to write
the computation explicitly, we will use the following notation:
\[
\hat B_{\k}^{+} \to \hat B_{\k}^{+}(\hat\mm), \qquad
\hat A_{\k}^{+} \to \hat A_{\k}^{+}(\gamma\hat\mm)\, .
\]
Then, the symmetry operators read
\be
\begin{array}{l}
\hat X_\k^{+}=  \hat A_{\k}^{+}\bigl(\gamma(\hat\mm-n\hbar\k)+m\hbar\k \bigr)
\, \dots \,
\hat A_{\k}^{+}\bigl(\gamma(\hat\mm-n\hbar\k)+\hbar\k \bigr)\,
\hat B_{\k}^{+}(\hat\mm)
\, \dots \,
\hat B_{\k}^{+}(\hat\mm) ,
\\[2.ex]
\hat X_\k^{-}
=  \hat A_{\k}^{-} \bigl(\gamma(\hat\mm+n\hbar\k)-(m-1)\hbar\k \bigr)
\,\dots\,
\hat A_{\k}^{-} \bigl(\gamma(\hat\mm+n\hbar\k) \bigr)\,
\hat B_{\k}^{-}(\hat\mm)
\,\dots\,
\hat B_{\k}^{-}(\hat\mm) ,
\nonumber
\end{array}
\ee
where these operators are always assumed to act on the space of eigenfunctions 
$\Psi_\k(\xi,y)=\Xi_\k^\mmb(\xi)\,Y_\k^{\gamma\mmb}(y)$ of $\hat H^\xi_\k$ with
eigenvalues given by (\ref{eve3}) and (\ref{eve4}).

%%%%%%%%%%%%%%%%%%%%%%%%%%%%%%%%%%%%%%%%%%%%%%%%%%%%%%%%%

 \section*{{Acknowledgments}}

This work was partially supported by the Spanish Ministerio de Econom\1a y Competitividad    (MINECO) under projects MTM2013-43820-P and MTM2014-57129-C2-1-P, and by the Spanish Junta de Castilla y Le\'on  under grant BU278U14.     \c{S}.~Kuru acknowledges the warm hospitality at the  Department of
Theoretical Physics, University of Valladolid, Spain.

%%%%%%%%%%%%%%%%%%%%%%%%%%%%%%%%%%%%%%%%%%%%%%

%%%%%%%%%%%%%%%%%%%%%%%%%%%%%%

\end{document}